\documentclass[journal]{IEEEtran}
\usepackage{amsmath,amsfonts}
\usepackage{algorithmic}
\usepackage{algorithm}
\usepackage{array}
\usepackage{textcomp}
\usepackage{stfloats}
\usepackage{url}
\usepackage{verbatim}
\usepackage{graphicx}
\usepackage{cite}
\usepackage{subcaption}

\begin{document}

\title{\huge Baseband-Free End-to-End Communication System Based on Diffractive Deep Neural Network}

\author{Xiaokun Teng,\IEEEmembership{}
	Wankai Tang,\IEEEmembership{~Member,~IEEE,}
	Xiao Li,\IEEEmembership{~Member,~IEEE,}
	and Shi Jin,\IEEEmembership{~Fellow,~IEEE}
	
	\thanks{Xiaokun Teng, Wankai Tang, Xiao Li, and Shi Jin are with the National Mobile Communications Research Laboratory, Southeast University, Nanjing 210096, China (e-mail: xkteng@seu.edu.cn; tangwk@seu.edu.cn; li\_xiao@seu.edu.cn; jinshi@seu.edu.cn).}
	
}

\markboth{Journal of \LaTeX\ Class Files,~Vol.~14, No.~8, August~2021}%
{Shell \MakeLowercase{\textit{et al.}}: A Sample Article Using IEEEtran.cls for IEEE Journals}


\maketitle

\begin{abstract}
Diffractive deep neural network (D\textsuperscript{2}NN), also referred to as reconfigurable intelligent metasurface based deep neural networks (Rb-DNNs) or stacked intelligent metasurfaces (SIMs) in the field of wireless communications, has emerged as a promising signal processing paradigm that enables computing-by-propagation. However, existing architectures are limited to implementing specific functions such as precoding and combining, while still relying on digital baseband modules for other essential tasks like modulation and detection. In this work, we propose a baseband-free end-to-end (BBF-E2E) wireless communication system where modulation, beamforming, and detection are jointly realized through the propagation of electromagnetic (EM) waves. The BBF-E2E system employs D\textsuperscript{2}NNs at both the transmitter and the receiver, forming an autoencoder architecture optimized as a complex-valued neural network. The transmission coefficients of each metasurface layer are trained using the mini-batch stochastic gradient descent method to minimize the cross-entropy loss. To reduce computational complexity during diffraction calculation, the angular spectrum method (ASM) is adopted in place of the Rayleigh–Sommerfeld formula. Extensive simulations demonstrate that BBF-E2E achieves robust symbol transmission under challenging channel conditions with significantly reduced hardware requirements. In particular, the proposed system matches the performance of a conventional multi-antenna system with 81 RF chains while requiring only a single RF chain and 1024 passive elements of metasurfaces. These results highlight the potential of wave-domain neural computing to replace digital baseband modules in future wireless transceivers.
\end{abstract}

\begin{IEEEkeywords}
Diffractive neural network, end-to-end communication, baseband-free, angular spectrum method, reconfigurable intelligent metasurface (RIS).
\end{IEEEkeywords}

\section{Introduction}
\subsection{Background}
\IEEEPARstart{E}{fficient} utilization of spatial resources has long been a key driver for the development of wireless communication technologies. Since the pioneering work of Foschini et al. \cite{foschiniLimitsWirelessCommunications1998}, multiple-input multiple-output (MIMO) technology has been extensively studied, leading to significant improvements in system capacity and spectral efficiency. Toward higher diversity and multiplexing gain, the concept of massive MIMO was introduced by further scaling up the number of antenna elements \cite{marzettaNoncooperativeCellularWireless2010}. However, increasing the number of antennas also introduces significant challenges, including higher costs, power consumption, and system complexity. Against this backdrop, reconfigurable intelligent surface (RIS) has emerged as one of the most promising technologies for future wireless communication systems, garnering substantial research attention in recent years for its ability to reconfigure wireless channels with low cost and power consumption \cite{direnzoSmartRadioEnvironments2020, wuIntelligentReflectingSurfaceaided2021, tangWirelessCommunicationsReconfigurable2020, huangReconfigurableIntelligentSurfaces2019, hanLargeIntelligentSurfaceassisted2019}. This indicates a paradigm shift in spatial resource exploitation, transitioning from merely increasing the number of active antenna elements to designing novel antenna arrays with enhanced structures and functionalities \cite{muSimultaneouslyTransmittingReflecting2022, zhangActiveRISVs2023, wuWidebandAmplifyingFiltering2024, yangIntelligentAdaptiveMetasurface2024}.

As a counterpart of conventional reflective RIS, transmissive RIS (T-RIS) enables simultaneous signal manipulation and transmission \cite{tangWirelessCommunicationsProgrammable2020, liTransmissiveRISTransceiver2024}, thus can be spatially cascaded to form multi-layer interconnected structures. Leveraging this property, a highly reconfigurable wave-domain signal processing architecture was proposed, referred to as a diffractive deep neural network (D\textsuperscript{2}NN) \cite{linAllopticalMachineLearning2018}. By jointly optimizing the transmission coefficients of each T-RIS layer, EM waves can be sequentially processed during propagating in D\textsuperscript{2}NN, realizing ``computing-by-propagation''. In the wireless communication research community, D\textsuperscript{2}NN has evolved into architectures like RIS based deep neural networks (Rb-DNNs) \cite{wangInterplayRISAI2021} and stacked intelligent metasurfaces (SIM) \cite{anStackedIntelligentMetasurfaces2023a}. These novel computational architectures not only overcome the limitation in conventional wireless communication systems, where signal processing is confined at digital baseband, but also facilitate the deep integration of wireless communications and artificial intelligence (AI), thanks to its intrinsic structural resemblance to neural networks.

\subsection{Related Work}

\subsubsection{D\textsuperscript{2}NN}

As Moore’s law approaches its physical limits, the advancement of integrated circuit technology is expected to slow down in the foreseeable future, making it increasingly challenging to meet the rapidly expanding computational demands \cite{gholamiAIMemoryWall2024}. In response to this challenge, Lin et al. proposed the D\textsuperscript{2}NN \cite{linAllopticalMachineLearning2018}. D\textsuperscript{2}NN is an optical diffractive computing architecture inspired by diffraction phenomena, potentially offering higher energy efficiency and lower latency than its electronic counterparts. In D\textsuperscript{2}NN, multiple transmissive metasurfaces with specific transmission responses are stacked over a relatively short distance. As EM waves propagate through these layers, the controlled diffraction and transmission enable high-speed signal processing. Inspired by D\textsuperscript{2}NN, numerous similar diffractive computing architectures have emerged \cite{zhouLargescaleNeuromorphicOptoelectronic2021, liuProgrammableDiffractiveDeep2022, chenAllanalogPhotoelectronicChip2023, gaoSuperresolutionDiffractiveNeural2024, guClassificationMetalHandwritten2024, liuDiffractiveDeepNeural2023, wetzsteinInferenceArtificialIntelligence2020, chenDiffractiveDeepNeural2024}. In \cite{zhouLargescaleNeuromorphicOptoelectronic2021}, a reconfigurable diffractive processing unit was constructed using optical devices like spatial light modulator, which can be programmed to implement various neural network architectures. In the millimeter-wave band , Liu et al. \cite{liuProgrammableDiffractiveDeep2022} developed a programmable D\textsuperscript{2}NN based on T-RIS hardware, demonstrating its image recognition and classification capabilities in related tasks. A multi-user transmission experiment was also conducted in \cite{liuProgrammableDiffractiveDeep2022}, highlighting the potential of D\textsuperscript{2}NN to enhance wireless communication systems. Another notable advancement beyond D\textsuperscript{2}NN is the development of optoelectronic-integrated neural networks \cite{chenAllanalogPhotoelectronicChip2023}, which leverage fully analog computation to eliminate dependence on digital signal processing, achieving higher energy efficiency and lower latency compared to state-of-the-art electronic processors. Additionally, D\textsuperscript{2}NN has been applied to super-resolution direction-of-arrival (DoA) estimation \cite{gaoSuperresolutionDiffractiveNeural2024}, handwritten digit recognition \cite{guClassificationMetalHandwritten2024}, and microwave imaging \cite{liuDiffractiveDeepNeural2023}, demonstrating its ability to perform various signal processing tasks.

\subsubsection{SIM}

Driven by the significant potential of D\textsuperscript{2}NN for wave-domain signal processing, a novel technology called SIM was recently proposed in the wireless communication research community \cite{anStackedIntelligentMetasurfaces2023a}. SIM essentially shares the same architecture as D\textsuperscript{2}NN. Therefore, SIM inherits the capability of ``computing-by-propagation'', considered as a low-cost, energy-efficient alternative to digital signal processing. A number of research has been conducted to explore the applications of SIM for beamforming, direction-of-arrival (DoA) estimation, integrated sensing and communication (ISAC), and semantic communications \cite{hassanEfficientBeamformingRadiation2024, anStackedIntelligentMetasurfaces2023c, anTwoDimensionalDirectionofArrivalEstimation2024, wangMultiuserISACStacked2024, yaoChannelEstimationStacked2024, huangStackedIntelligentMetasurfaces2025}. In these studies, increasing the number of SIM layers has demonstrated remarkable performance gains, highlighting the advantages of SIM over a single-layer T-RIS. For instance, compared to a single-layer T-RIS, a three-layer SIM can improve the beam power concentration in a designated region by 50\% \cite{hassanEfficientBeamformingRadiation2024}. Meanwhile, increasing the number of SIM layers helps reduce the mean squared error (MSE) in matrix fitting problems. In \cite{anStackedIntelligentMetasurfaces2023a}, SIM was used to approximate the precoding and combining matrices and replace the corresponding modules in the digital baseband of holographic MIMO communication systems. Similarly, \cite{anTwoDimensionalDirectionofArrivalEstimation2024} employed SIM to approximate the discrete fourier transform (DFT) matrix for DoA estimation and proposed a protocol that enhances estimation precision through input-layer design. Experimentally, a recent study \cite{wangMultiuserISACStacked2024} implemented a SIM-aided communication and sensing prototype system, which demonstrated the capability of SIM for both wireless link enhancement and dynamic DoA estimation. Despite these advancements, the theoretical and experimental research of SIM is still in its early stages, leaving many open challenges to be addressed.

\subsubsection{E2E communication}

Existing research on SIM has primarily focused on replacing individual modules within wireless communication systems, such as the precoding and combining modules \cite{anStackedIntelligentMetasurfaces2023a}. However, the digital baseband is still responsible for essential tasks such as modulation and detection, preventing SIM from fully eliminating the reliance on digital signal processing. To address this limitation, incorporating SIM into an end-to-end (E2E) communication framework offers a promising pathway to extending its capability. E2E communication architectures typically employ auto-encoders (AEs) to unify multiple communication modules into a single framework. With the powerful capabilities of deep learning, an AE can autonomously learn the mapping from bit streams to transmitted symbols, forming an encoding network, while the receiver reconstructs the transmitted symbols using a decoding network \cite{osheaIntroductionDeepLearning2017, dornerDeepLearningBased2018}. Unlike the traditional modular systems, E2E communication systems follow a data-driven, globally optimized design paradigm, offering performance improvements beyond conventional architectures. Over the past few years, substantial research has explored the performance of E2E communication systems under various antenna configurations and channel conditions \cite{qinAIEmpoweredWireless2024}. For instance, in \cite{yeDeepLearningBased2021}, a pilot-free E2E wireless communication scheme was developed for flat-fading MIMO channels. Instead of relying on pilot signals for channel estimation, the receiver was designed with two deep neural network modules to infer channel characteristics and recover transmitted data. Similarly, \cite{aitaoudiaEndtoEndLearningOFDM2022} investigated E2E learning over frequency- and time-selective fading channels using OFDM waveforms, demonstrating performance gains achieved through pilot and constellation learning. Different from other research that assume known and differentiable channel models, \cite{yeDeepLearningBasedEndtoEnd2020} introduced a conditional generative adversarial network to model channel effects, enabling a differentiable connection between the transmitter and receiver.

\subsection{Motivation and Contribution}
Owing to the high flexibility and maturity of digital circuits, shifting RF signals to the digital baseband for processing and computation has long been the prevailing paradigm in wireless communications. However, the emergence of D\textsuperscript{2}NN and SIM has begun to challenge this conventional paradigm. With superior energy efficiency and lower latency, direct signal processing in the wave domain holds the potential to offload computationally intensive tasks from the digital baseband, thereby reducing power consumption and processing delays. However, the current research has been limited to using SIM to replace an individual module in communication systems, while the digital baseband remains responsible for other signal processing tasks such as modulation and detection. Consequently, these studies have yet to fully eliminate the dependence of wireless communication systems on synchronous transceiver and digital baseband. To bridge this gap, we design a D\textsuperscript{2}NN-enabled architecture that establishes a baseband-free wireless communication scheme in an E2E manner. The primary contributions of this work are summarized as follows:

\begin{itemize}
	\item{We propose a baseband-free end-to-end (BBF-E2E) communication system based on D\textsuperscript{2}NN. In this system, both the transmitter and receiver are equipped with D\textsuperscript{2}NNs, enabling direct wave-domain signal processing and forming an AE architecture. The D\textsuperscript{2}NNs at the transceiver function as the encoder and decoder, respectively. This communication scheme completely eliminates the need for digital baseband modules, achieving a baseband-free communication paradigm.}
	\item{We develop the equivalent neural network expression and the training framework for BBF-E2E, where the cross-entropy (CE) loss serves as the optimization metric. The phase shift parameters of each metasurface layer in D\textsuperscript{2}NNs are optimized to minimize the loss function through the mini-batch stochastic gradient descent (SGD) algorithm. The gradients of the loss function with respect to the D\textsuperscript{2}NN parameters are derived through Wirtinger calculus. }
	\item{We introduce the angular spectrum method (ASM) for efficient training of D\textsuperscript{2}NNs. ASM transforms the spatial domain signal into the frequency-domain to accelerate diffraction calculation. Our proposed method leverages the fast Fourier transform (FFT) to reduce the computational complexity from $O(N^{2})$ to $O(N\log N)$ for diffraction computations between adjacent layers. Furthermore, zero-padding is applied to mitigate the issue of circular convolution, thereby improving the accuracy of ASM.}
	\item{Extensive simulation results demonstrate the performance of the BBF-E2E system under various system configurations and channel conditions. As the number of layers and elements in D\textsuperscript{2}NN increases, both training loss and symbol error rate (SER) performance improve. In performance comparison, the proposed BBF-E2E system achieves the same performance as a conventional beamforming and modulation scheme that requires 81 RF chains, yet it attains this with only a single RF chain and 1024 metasurface elements.}
\end{itemize}

\begin{figure*}[t]
	\centering
	\includegraphics[width=0.9\textwidth]{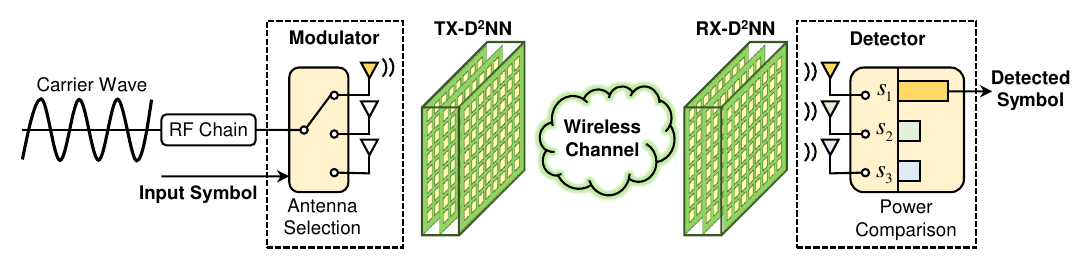}
	\vspace{-0.1cm}
	\caption{Transceiver Architecture of the BBF-E2E system. The input symbol is mapped to a selected antenna subarray by the modulator, from which a pure carrier wave is radiated. The TX-D\textsuperscript{2}NN encodes the input field for robust transmission through the wireless channel. The RX-D\textsuperscript{2}NN decodes the received signal and concentrates signal power within a specific region for power-based detection.}
	\label{fig:transceiverArchitecture}
	\vspace{-0.4cm}
\end{figure*}

\subsection{Organization and Notations}
The remainder of this paper is organized as follows. Section II introduces the proposed BBF-E2E transceiver architecture, outlining its key components. Section III presents the system model of BBF-E2E, highlighting the process of wave-domain signal processing. Section IV details the forward and backward propagation algorithms for efficient BBF-E2E training. Section V provides numerical results and analysis. Finally, Section VI concludes the paper and discusses potential directions for future research.

\textit{Notations}: Throughout this paper, we adopt the following notations. The imaginary unit is denoted by $j$. Boldface lowercase and uppercase letters represent vectors and matrices, respectively. For a vector $\mathbf{v}$, both $\mathbf{v}[m]$ and $v_m$ denote its $m$-th element. Similarly, for a matrix $\mathbf{V}$, both $\mathbf{V}[m_x,m_z]$ and $v_{m_x,m_z}$ represent the element at row $m_x$ and column $m_z$. The submatrix extraction operation is expressed as $\mathbf{V}_{\text{sub}}=\mathbf{V}[m_{x}:m_{x}',m_{z}:m_{z}']$. The transpose and conjugate transpose of a matrix are denoted by $\left( \cdot \right)^{\mathrm{T}}$ and $\left( \cdot \right)^{\mathrm{H}}$, respectively. The operation $\mod(x,y)$ returns the remainder when $x$ is divided by $y$. The notation $\operatorname{diag}(\mathbf{x})$ represents a diagonal matrix whose diagonal elements are taken from the vector $\mathbf{x}$. The vectorization operator $\operatorname{vec}(\cdot)$ stacks the columns of a matrix into a column vector. The symbol $\odot$ and $\otimes$ denotes the Hadamard product and the Kronecker product, respectively.

\section{Transceiver Architecture}
\label{transceiverArchitecture}

BBF-E2E performs a direct mapping from input symbols to detected symbols without the need for digital baseband. To achieve this, BBF-E2E transceiver consists of five key components: a modulator, a transmitter-side (TX) D\textsuperscript{2}NN, the wireless channel, a receiver-side (RX) D\textsuperscript{2}NN, and a detector. The overall architecture of BBF-E2E system is illustrated in Fig.~\ref{fig:transceiverArchitecture}, and each component is detailed below.

\subsubsection{Modulator}
The modulator comprises an antenna selection module and multiple feed antennas. The antenna selection module maps the input symbol to a specific antenna subarray index, activating one set of feed antennas to radiate the EM wave while keeping all others silent. A pure carrier wave is radiated by the activated feed antennas, therefore the transmitter only needs a single RF chain. The mapping from the input symbol to the activated feed antennas can be flexibly designed. More than one antenna can be simultaneously activated, which slightly differs from the conceptual depiction in Fig.~\ref{fig:transceiverArchitecture}.

\subsubsection{TX-D\textsuperscript{2}NN}
The TX-D\textsuperscript{2}NN consists of $L_{\text{TX}}$ cascaded T-RIS layers, forming a wave-domain signal processing unit based on the D\textsuperscript{2}NN architecture. The input field generated by the modulator propagates through multiple metasurface layers, where it undergoes wave-domain signal processing before being radiated into the wireless channel. Each element of these layers can independently adjust the phase of the transmissive signal, enabling TX-D\textsuperscript{2}NN to encode the input field into a structured radiation field. 

\subsubsection{Wireless Channel}
The wireless channel introduces stochastic noise and multipath fading that affect signal transmission. In this study, we adopt correlated Rician fading channels to evaluate the robustness of BBF-E2E against noise and fading distortions, while considering the correlation introduced by the close spacing of metasurface elements.

\subsubsection{RX-D\textsuperscript{2}NN}
The RX-D\textsuperscript{2}NN is structurally similar to TX-D\textsuperscript{2}NN, consisting of $L_{\text{RX}}$ stacked T-RIS layers. It serves as a wave-domain decoder, demodulating the noisy and faded EM signal and facilitating efficient detection. After wave-domain processing in RX-D\textsuperscript{2}NN, the energy of the output field is expected to concentrate within a specific region, allowing the detector to recover the transmitted symbol with minimal complexity.

\subsubsection{Detector}
The detector consists of receiving antennas and a power comparison module, converting the output field into the detected symbol. In the power comparison module, there are power detectors whose outputs represent the probability of the corresponding symbol being transmitted. The detected symbol is determined by identifying the power detector with the highest output and mapping its index to the corresponding symbol.

\newtheorem{remark}{Remark}

\begin{remark}
BBF-E2E enables direct wave-domain signal processing, eliminating the need for digital baseband and synchronous transceiver. The transmitter requires only a single RF chain, while the receiver operates using low-cost power detectors, significantly reducing hardware cost. Since symbol decisions are made based on power detection, BBF-E2E operates in a non-coherent mode without requiring strict carrier phase synchronization during transmission. Overall, BBF-E2E exploits the strong capability of D\textsuperscript{2}NN, embedding modulation and beamforming operations into the propagation of EM waves at the speed of light. This ``computing-by-propagation'' paradigm holds the potential to drastically reduce system complexity and redefine the architecture of wireless communication systems.
\end{remark}

\begin{figure*}[ht!]
	\centering
	\begin{subfigure}{0.48\textwidth}
		\includegraphics[width=\linewidth]{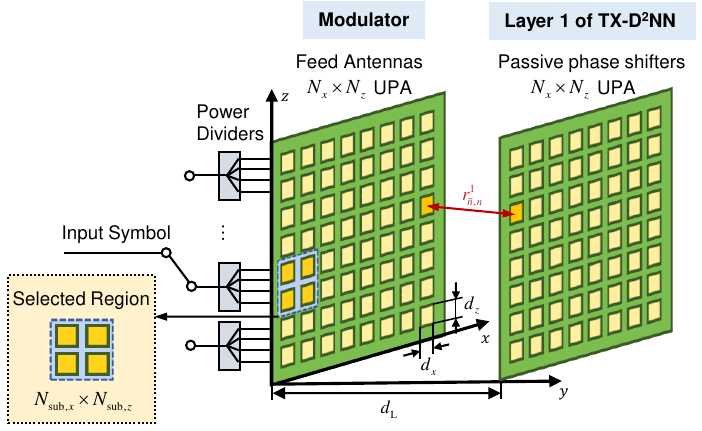}
		\caption{}
		\label{fig:modelsub1}
	\end{subfigure}\hfill
	\begin{subfigure}{0.48\textwidth}
		\includegraphics[width=\linewidth]{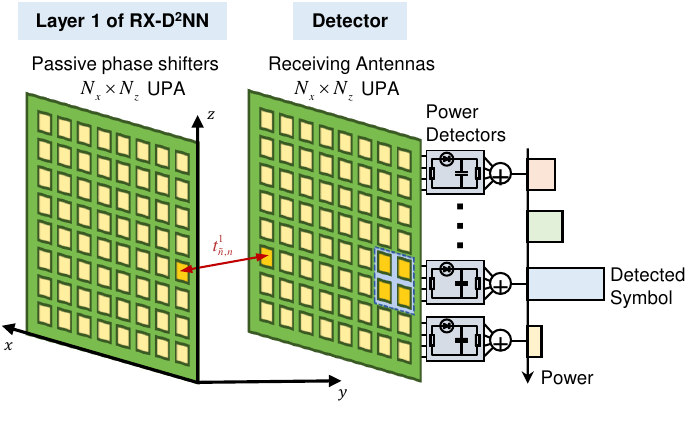}
		\caption{}
		\label{fig:modelsub2}
	\end{subfigure}
	\vspace{-0.1cm}
	\caption{Detailed illustration of partial components of BBF-E2E. (a) The modulator and layer 1 of TX-D\textsuperscript{2}NN. (b) The detector and layer 1 of RX-D\textsuperscript{2}NN.}
	\label{fig:systemModel}
	\vspace{-0.4cm}
\end{figure*}

\section{System Model}
\label{systemModel}

In this section, we present a detailed model to characterize the BBF-E2E system. For convenience, we assume that the feed antennas, all D\textsuperscript{2}NN layers, and receiving antennas share the same uniform planar array (UPA) structure, as shown in Fig.~\ref{fig:systemModel}. Each layer consists of $N_x$ columns along the $x$-axis and $N_z$ rows along the $z$-axis, resulting in a total of $N = N_xN_z$ elements. Therefore, the diffraction between the feed or receiving antennas and the D\textsuperscript{2}NN follows the same principles as intra-D\textsuperscript{2}NN diffraction, enabling unified diffraction modeling. For convenience, we denote $\mathcal{N}=\left\{ 1,2,\dots,N \right\}$, $\mathcal{L}_\text{TX}=\left\{ 1,2,\dots,L_\text{TX} \right\}$, $\mathcal{L}_\text{RX}=\left\{ 1,2,\dots,L_\text{RX} \right\}$

Let \( \mathcal{S} = \{s_1, s_2, \dots, s_M\}\) and $\mathcal{M}=\{1,2,\dots,M\}$, where $\mathcal{S}$ denotes the symbol set, and $M$ represents the modulation order, indicating the number of antenna subarrays. We assume $M = 2^p$ and $p \in \mathbb{N}^{+}$ is a positive even integer, representing the number of transmitted bits per symbol. For each symbol $s \in \mathcal{S}$, we map $s$ uniquely to a one-hot vector $\mathbf{q}_s \in \{0, 1\}^M$ to determine which feed antenna subarray is activated, where $\forall m \in \mathcal{M}$, we have
\begin{align}
	\mathbf{q}_s[m] = 
	\begin{cases} 
		1, & \text{if } s = s_m, \\
		0, & \text{otherwise}.
	\end{cases}
\end{align}

Next, we describe the mapping process from the transmitted symbol $s_{m}$ to the signal transmitted from the modulator. As illustrated in Fig.~\ref{fig:systemModel}(a), each transmitting symbol corresponds to the activation of a specific feed antenna subarray. To formulate this, let $M = M_x M_z$, where $M_x$ and $M_z$ denote the modulation order along the $x$-axis and $z$-axis, respectively. We assume that $N_x$ and $N_z$ are divisible by $M_x$ and $M_z$, respectively. Define $N_{\text{sub},x} = N_x / M_x$ and $N_{\text{sub},z} = N_z / M_z$, such that $N_{\text{sub}} = N_{\text{sub},x} N_{\text{sub},z}$ represents the number of sub-array elements activated per transmission. To indicate the location of the activated sub-array, we introduce the one-hot matrix $\mathbf{Q}_s \in \{0,1\}^{M_x \times M_z}$, where
\begin{align}
	\mathbf{Q}_s[m_x, m_z] = 
	\begin{cases} 
		1, & \text{if } s = s_{m_z M_x + m_x}, \\
		0, & \text{otherwise},
	\end{cases}
\end{align}
which satisfies $\mathbf{q}_s = \text{vec}(\mathbf{Q}_s)$. Thus, the signal transmitted from the modulator, denoted as $\mathbf{U}_0 \in \mathbb{C}^{N_x \times N_z}$, is given by:
\begin{align}
	\label{eq:transmittedSignal}
	\mathbf{U}_0 = \frac{1}{\sqrt{N_\text{sub}}}\mathbf{Q}_s \otimes \mathbf{1}^{N_{\text{sub},x} \times N_{\text{sub},z}},
\end{align}
since all activated feed antennas are connected to the same RF chain through power dividers and transmit pure carriers. For instance, Fig.~\ref{fig:systemModel}(\subref{fig:modelsub1}) depicts a modulator with $N_x=N_z=8$ and $M_x=M_z=4$. The highlighted feed antennas are selected and the transmitted signal can be express with \eqref{eq:transmittedSignal}.

For TX-D\textsuperscript{2}NN, the spacing between adjacent metasurface layers is uniform and denoted as $d_\text{L}$. Each element in all layers can independently adjust the phase of the transmitted signal. Let $d_x$ and $d_z$ represent the element dimensions along the $x$- and $z$-axes, respectively. The area of each element is then given by $A = d_x d_z$. Denoting the transmission phase of the $n$-th element in the $l$-th layer of TX-D\textsuperscript{2}NN as $\beta_{l,n}$, the transmission coefficient matrix of the $l$-th metasurface layer is expressed as $\mathbf{\Phi}^{l} = \text{diag}\left(\left[ \phi_{l,1},\phi_{l,2},\dots ,\phi_{l,N} \right]^{\mathrm{T}} \right)$, where $\phi_{l,n} = e^{j\beta_{l,n}}$, $\forall l \in \mathcal{L}_\text{TX}$ and $n \in \mathcal{N}$.

The diffractive propagation between metasurface layers is modeled using Rayleigh–Sommerfeld diffraction theory~\cite{goodmanIntroductionFourierOptics2005}. Specifically, let $\mathbf{W}^{l} = \left( w_{n,\tilde{n}}^{l} \right)_{N\times N}$ represent the propagation coefficient matrix between the $(l-1)$-th and $l$-th metasurface layers in TX-D\textsuperscript{2}NN, where the feed antennas is regarded as the 0-th layer of TX-D\textsuperscript{2}NN. The propagation coefficient between the $n$-th element of the $(l-1)$-th layer and the $\tilde{n}$-th element of the $l$-th layer is given by
\begin{equation}
	\begin{aligned}
		w^{l}_{\tilde{n},n}=\frac{Ad_\text{L}}{\left( r_{\tilde{n},n}^{l} \right)^{2}}\left( \frac{1}{2\pi r_{\tilde{n},n}^{l}}+\frac{1}{j\lambda} \right)e^{j\frac{2\pi r_{\tilde{n},n}^{l}}{\lambda}},
	\end{aligned}
\end{equation}
for $\forall l \in \mathcal{L}_\text{TX}, n \in \mathcal{N}$ and $\tilde{n} \in \mathcal{N}$, where $r_{\tilde{n},n}^{l}$ denotes the distance between these two elements, and $\lambda$ represents the electromagnetic wavelength. The term $\frac{1}{2\pi r_{\tilde{n},n}^{l}}$ accounts for evanescent waves, which decay rapidly within a few wavelengths but must be considered in stacked metasurface structures due to the small inter-layer spacing. The distance $r_{\tilde{n},n}^{l}$ is given by $r_{\tilde{n},n}^{l}=\sqrt{ d_\text{L}^{2}+(n_{x}-\tilde{n}_{x})^{2}d_{x}^{2}+(n_{z}-\tilde{n}_{z})^{2}d_{z}^{2} }$, $\forall l \in \mathcal{L}_\text{TX}, n \in \mathcal{N}$ and $\tilde{n} \in \mathcal{N}$, where $n_x$ and $n_z$ denote the indices of the $n$-th element along the $x$- and $z$-axes, respectively, satisfying $n = n_z N_x + n_x$. $\tilde{n}_x$ and $\tilde{n}_z$ are defined analogously for the $\tilde{n}$-th element.

Let $\mathbf{u}_{l}$ denote the signal radiated from the $l$-th layer of TX-D\textsuperscript{2}NN, $\forall l \in \mathcal{L}_\text{TX}$. Then, the processing by TX-D\textsuperscript{2}NN on the input signal can be expressed as
\begin{align}
	\label{eq:TXNNSignalProcessing}
	\mathbf{u}_{L_\text{TX}} = \mathbf{\Phi}^{L_\text{TX}}\mathbf{W}^{L_\text{TX}} \dots \mathbf{\Phi}^{2}\mathbf{W}^{2}\mathbf{\Phi}^{1}\mathbf{W}^{1}\mathbf{u}_{0},
\end{align}
where $\mathbf{u}_{0}=\text{vec}(\mathbf{U}_0)$ and $\mathbf{u}_{L_\text{TX}}$ represents the signal radiated into the free space by the outermost layer of TX-D\textsuperscript{2}NN. By defining $\mathbf{B}_\text{TX} = \mathbf{\Phi}^{L_\text{TX}}\mathbf{W}^{L_\text{TX}} \dots \mathbf{\Phi}^{2}\mathbf{W}^{2}\mathbf{\Phi}^{1}\mathbf{W}^{1}$, \eqref{eq:TXNNSignalProcessing} can be further expressed as $\mathbf{u}_{L_\text{TX}} = \mathbf{B}_\text{TX} \mathbf{u}_{0}$.

After propagating through the wireless channel, the signal received by the outermost layer of RX-D\textsuperscript{2}NN $\mathbf{v}_{L_\text{RX}} \in \mathbb{C}^{N}$ can be expressed as
\begin{align}
	\mathbf{v}_{L_\text{RX}} = \mathbf{H} \mathbf{u}_{L_\text{TX}} + \mathbf{n},
\end{align}
where $\mathbf{H} \in \mathbb{C}^{N \times N}$ denotes the channel matrix, and $\mathbf{n} \sim \mathcal{CN}(\mathbf{0}, \sigma^2 \mathbf{I}_N)$ represents the additive white Gaussian noise, and $\sigma$ is the noise variance. In our work, we consider a correlated Rician fading channel that captures both deterministic line-of-sight (LoS) propagation and stochastic multipath effects. Under this channel model, $\mathbf{H}$ is modeled as a combination of the deterministic LoS component $\mathbf{H}_\text{LoS}$ and the stochastic non-line-of-sight (NLoS) component $\mathbf{H}_\text{NLoS}$:
\begin{align}
	\label{eq:channelModel}
	\mathbf{H} = \sqrt{ \frac{K_\text{R}}{1 + K_\text{R}} } \mathbf{H}_\text{LoS} + \sqrt{ \frac{1}{1 + K_\text{R}} } \mathbf{H}_\text{NLoS},
\end{align}
where $K_\text{R}$ is the Rician $K$-factor, indicating the power ratio between the LoS and NLoS components. The LoS channel component is modeled as
\begin{align}
	\mathbf{H}_\text{LoS} = \mathbf{a}_\text{TX}(\theta^{\text{ele}}_\text{TX}, \theta^{\text{azi}}_\text{TX}) \, \mathbf{a}_\text{RX}^{\mathrm{H}}(\theta^{\text{ele}}_\text{RX}, \theta^{\text{azi}}_\text{RX}),
\end{align}
where $\mathbf{a}_\text{TX}(\cdot)$ and $\mathbf{a}_\text{RX}(\cdot)$ denote the array response vectors of the outermost layers of TX-D\textsuperscript{2}NN and RX-D\textsuperscript{2}NN. The parameters $\theta^{\text{ele}}_\text{TX}, \theta^{\text{azi}}_\text{TX}$ and $\theta^{\text{ele}}_\text{RX}, \theta^{\text{azi}}_\text{RX}$ represent the elevation and azimuth angles of departure and arrival in the LoS path, respectively. Given the UPA structure in Fig.~\ref{fig:systemModel}, the array response vectors can be factorized as
\begin{align}
		\mathbf{a}_\text{TX}(\theta^{\text{ele}}_\text{TX}, \theta^{\text{azi}}_\text{TX}) &= \mathbf{a}_{\text{TX},x}(\theta^{\text{ele}}_\text{TX}, \theta^{\text{azi}}_\text{TX}) \otimes \mathbf{a}_{\text{TX},z}(\theta^{\text{ele}}_\text{TX}), \\
		\mathbf{a}_\text{RX}(\theta^{\text{ele}}_\text{RX}, \theta^{\text{azi}}_\text{RX}) &= \mathbf{a}_{\text{RX},x}(\theta^{\text{ele}}_\text{RX}, \theta^{\text{azi}}_\text{RX}) \otimes \mathbf{a}_{\text{RX},z}(\theta^{\text{ele}}_\text{RX}),
\end{align}
where $\mathbf{a}_{\text{TX},x}(\theta^{\text{ele}}_{\text{TX}}, \theta^{\text{azi}}_\text{TX}) = [1, e^{j\omega_{\text{TX},x}}, \dots, e^{j(N_{x}-1)\omega_{\text{TX},x}}]^{\mathrm{T}}$ is the $x$-axis array response with spatial frequency $\omega_{{\text{TX},x}} = \frac{2\pi d_x}{\lambda} \sin \theta^{\text{ele}}_\text{TX} \cos \theta^{\text{azi}}_\text{TX}$, and $\mathbf{a}_{\text{TX},z}(\theta^{\text{ele}}_{\text{TX}}) = [1, e^{j\omega_{\text{TX},z}}, \dots, e^{j(N_{z}-1)\omega_{\text{TX},z}}]^{\mathrm{T}}$ is the $z$-axis array response with spatial frequency $\omega_{{\text{TX},z}} = \frac{2\pi d_z}{\lambda} \cos \theta^{\text{ele}}_\text{TX}$. The expressions for $\mathbf{a}_{\text{RX},x}$ and $\mathbf{a}_{\text{RX},z}$ follow analogously by replacing the subscript "TX" with "RX". For the NLoS component, correlated Rayleigh fading is adopted, and the channel matrix is modeled as
\begin{align}
	\mathbf{H}_\text{NLoS} = \alpha \mathbf{R}_\text{RX}^{1/2} \mathbf{G} \mathbf{R}_\text{TX}^{1/2},
\end{align}
where the normalization factor $\alpha = N / \Vert \mathbf{R}_\text{RX}^{1/2} \mathbf{G} \mathbf{R}_\text{TX}^{1/2} \Vert_{\text{F}}$ ensures that $\Vert \mathbf{H}_\text{NLoS} \Vert_\text{F}^2 = N^2$~\cite{heathjrFoundationsMIMOCommunication2018}. Here, $\mathbf{G} \in \mathbb{C}^{N \times N}$ is an i.i.d. Rayleigh fading matrix with $\mathbf{G} \sim \mathcal{CN}(\mathbf{0}, \mathbf{I}_N \otimes \mathbf{I}_N)$. $\mathbf{R}_\text{TX}, \mathbf{R}_\text{RX} \in \mathbb{C}^{N \times N}$ denote the transmit and receive spatial correlation matrices, respectively, which is modeled by
\begin{align}
	\mathbf{R}_\text{TX}[n,\tilde{n}] = \mathbf{R}_\text{RX}[n,\tilde{n}] = \text{sinc}(2r_{n,\tilde{n}} / \lambda),
\end{align}
where $r_{n,\tilde{n}} = \sqrt{(n_x - \tilde{n}_x)^2 d_x^2 + (n_z - \tilde{n}_z)^2 d_z^2}$.

RX-D\textsuperscript{2}NN totally shares the same structure as that in TX-D\textsuperscript{2}NN. Let $\gamma_{l,n}$ denote the transmission phase of the $n$-th element in the $l$-th layer of RX-D\textsuperscript{2}NN. The transmission coefficient matrix of the $l$-th layer can be expressed as $\mathbf{\Psi}^{l} = \text{diag}\left(\left[ \psi_{l,1}, \psi_{l,2}, \dots, \psi_{l,N} \right]^{\mathrm{T}} \right)$, where $\psi_{l,n} = e^{j\gamma_{l,n}}$, $\forall l \in \mathcal{L}_\text{RX}$. Similar to TX-D\textsuperscript{2}NN, the diffractive propagation between metasurface layers of RX-D\textsuperscript{2}NN is modeled using the Rayleigh–Sommerfeld diffraction theory. Let $\mathbf{Z}^{l} = \left( z_{\tilde{n},n}^{l} \right)_{N \times N}$ denote the propagation matrix from the $l$-th layer to the $(l-1)$-th layer in RX-D\textsuperscript{2}NN, where the receiving antennas is regarded as the 0-th layer of RX-D\textsuperscript{2}NN. The propagation coefficient between the $n$-th element in the $l$-th layer and the $\tilde{n}$-th element in the $(l-1)$-th layer is given by
\begin{align}
	z^{l}_{\tilde{n},n} = \frac{A d_\text{L}}{\left( t_{\tilde{n},n}^{l} \right)^2} \left( \frac{1}{2\pi t_{\tilde{n},n}^{l}} + \frac{1}{j\lambda} \right) e^{j\frac{2\pi t_{\tilde{n},n}^{l}}{\lambda}},
\end{align}
for $\forall l \in \mathcal{L}_\text{RX}$, $n \in \mathcal{N}$ and $\tilde{n} \in \mathcal{N}$, where $t_{\tilde{n},n}^{l}$ denotes the distance between the two elements of adjacent layers, and follows the same expression as $r_{\tilde{n},n}^{l}$ defined previously.

Let $\mathbf{v}_{l}$ denote the signal received by the $l$-th layer of RX-D\textsuperscript{2}NN, $\forall l \in \mathcal{L}_\text{RX}$, and let $\mathbf{v}_0$ represent the output field formed on the detector array after processing through RX-D\textsuperscript{2}NN. Then, the overall signal process performed by RX-D\textsuperscript{2}NN can be expressed as
\begin{align}
	\label{eq:RXNNSignalProcessing}
	\mathbf{v}_0 = \mathbf{Z}^{1} \mathbf{\Psi}^{1} \mathbf{Z}^{2} \mathbf{\Psi}^{2} \dots \mathbf{Z}^{L_\text{RX}} \mathbf{\Psi}^{L_\text{RX}} \mathbf{v}_{L_\text{RX}}.
\end{align}
Defining $\mathbf{B}_\text{RX} = \mathbf{Z}^{1} \mathbf{\Psi}^{1} \mathbf{Z}^{2} \mathbf{\Psi}^{2} \dots \mathbf{Z}^{L_\text{RX}} \mathbf{\Psi}^{L_\text{RX}}$, we obtain $\mathbf{v}_0 = \mathbf{B}_\text{RX} \mathbf{v}_{L_\text{RX}}.$

\begin{figure*}[t]
	\centering
	\includegraphics[width=\textwidth]{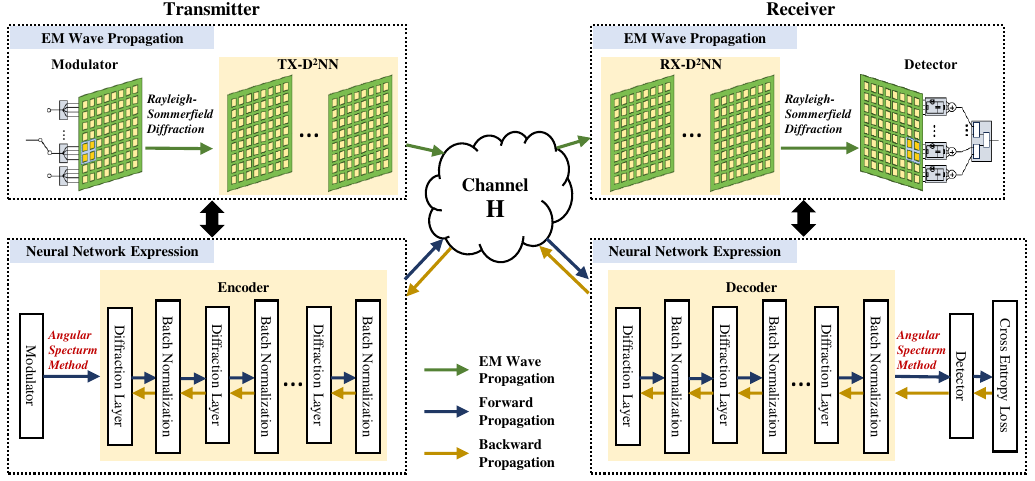}
	\vspace{-0.1cm}
	\caption{The learning framework of BBF-E2E system. The EM wave propagation in the upper half is abstracted as the neural network expression in the lower half.}
	\label{fig:ASMFramework}
	\vspace{-0.4cm}
\end{figure*}

In the detector, similar to the modulator, the $N$ antennas are uniformly divided into $M$ subarrays, each containing $N_\text{sub}$ elements, as illustrated in Fig.~\ref{fig:systemModel}(b). Let $\mathbf{V}_{0} \in \mathbb{C}^{N_x \times N_z}$ denotes the matrix representation of $\mathbf{v}_{0}$, i.e., $\mathbf{v}_{0}=\text{vec}(\mathbf{V}_{0})$. Use $\mathbf{V}_m$ to extract the received signal on the $m$-th sub-array, where
\begin{align}
	\mathbf{V}_m=\mathbf{V}_{0}\odot(\mathbf{Q}_{s_m} \otimes \mathbf{1}^{N_{\text{sub},x} \times N_{\text{sub},z}}), \forall m \in \mathcal{M}.
\end{align}
Denote $p_m$ as the total output power of the $m$-th sub-array, which reflects the probability that symbol $s_m$ has been transmitted. It is computed as
\begin{align}
	p_m = \mathrm{Tr}\left( \mathbf{V}_{m}^{\mathrm{H}} \mathbf{V}_{m} \right), \forall m \in \mathcal{M}.
\end{align}
Let $\mathbf{p} = [p_1, p_2, \dots, p_M]^{\mathrm{T}}$, thus the overall signal processing of the BBF-E2E system from the input one-hot vector $\mathbf{q}_s$ to the output $\mathbf{p}$ can then be represented as
\begin{align}
	\mathbf{p} &= f_\text{det}\left( \mathbf{B}_\text{RX} \mathbf{H} \mathbf{B}_\text{TX} f_\text{mod}(\mathbf{q}_s) + \mathbf{n}' \right),
\end{align}
where $\mathbf{n}' = \mathbf{B}_\text{RX} \mathbf{n}$, and $f_\text{mod}$ and $f_\text{det}$ denote the processing functions of the modulator and detector, respectively. Finally, the decision is made by selecting the sub-array with the maximum detected power:
\begin{align}
	\label{detector}
	\hat{m} = \mathop{\text{argmax}}\limits_{m \in \mathcal{M}} \left\{ p_m \right\},
\end{align}

\vspace{-0.2cm}
\section{Forward and Backward Propagation for Fast Training}
The BBF-E2E system is designed to mitigate the effects of channel noise and fading by leveraging a D\textsuperscript{2}NN-based autoencoder to recover the input one-hot vector $\mathbf{q}_s$ from the output $\mathbf{p}$. To avoid the non-differentiability issue caused by the hard maximization in~\eqref{detector}, we apply the softmax function to convert the output into a probability distribution. Specifically, define $\mathbf{p}' = \text{softmax}(\mathbf{p})$, where $\mathbf{p}'[m] = \frac{e^{p_m}}{\sum_{m=1}^{M} e^{p_m}}, \forall m \in \mathcal{M}.$ The CE loss is adopted to measure the divergence between $\mathbf{q}_s$ and $\mathbf{p}'$. Denoting the CE loss function as $\mathcal{L}_\text{CE}(\mathbf{q}_s, \mathbf{p}')$, it can be formulated as
\begin{equation}
	\label{eq:lossFunction}
	\mathcal{L}_\text{CE}(\mathbf{q}_{s}, \mathbf{p}') = \mathbb{E}_{s,\mathbf{n}} \left\{ -\sum_{m=1}^{M} q_{s,m} \log p_{m}' \right\}.
\end{equation}
Then the problem of BBF-E2E can be formulated as
\begin{equation}
	\begin{aligned}
		\text{(P1)}: \mathop{\min}\limits_{\mathbf{\Phi}^{l}, \mathbf{\Psi}^{l}} \quad & \mathcal{L}_\text{CE}(\mathbf{q}_s, \mathbf{p}') \\
		\text{s.t.} \quad & \mathbf{p}' = \text{softmax}(\mathbf{p}), \\
		& \mathbf{p} = f_\text{det} \left( \mathbf{B}_\text{RX} \mathbf{H} \mathbf{B}_\text{TX} f_\text{mod}(\mathbf{q}_s) + \mathbf{n}' \right), \\
		& \mathbf{B}_\text{TX} = \mathbf{\Phi}^{L_\text{TX}} \mathbf{W}^{L_\text{TX}} \dots \mathbf{\Phi}^{1} \mathbf{W}^{1}, \\
		& \mathbf{B}_\text{RX} = \mathbf{Z}^{1} \mathbf{\Psi}^{1} \mathbf{Z}^{2} \mathbf{\Psi}^{2} \dots \mathbf{Z}^{L_\text{RX}} \mathbf{\Psi}^{L_\text{RX}}, \\
		& \mathbf{\Phi}^{l} = \text{diag}\left( \left[ \phi_{l,1}, \phi_{l,2}, \dots, \phi_{l,N} \right]^\mathrm{T} \right), \forall l \in \mathcal{L}_\text{TX}, \\
		& \mathbf{\Psi}^{l} = \text{diag}\left( \left[ \psi_{l,1}, \psi_{l,2}, \dots, \psi_{l,N} \right]^\mathrm{T} \right), \forall l \in \mathcal{L}_\text{RX}, \\
		& |\phi_{l,n}| = 1, \quad \forall l \in \mathcal{L}_\text{TX}, \, n \in \mathcal{N}, \\
		& |\psi_{l,n}| = 1, \quad \forall l \in \mathcal{L}_\text{RX}, \, n \in \mathcal{N}.
	\end{aligned}
\end{equation}

Due to the strong coupling among optimization variables and the non-convex unit-modulus constraints on the transmission coefficients of metasurface elements, it is extremely challenging to obtain a global optimum of problem (P1). Notably, the structure of BBF-E2E closely resembles that of an autoencoder (AE). Hence, we leverage machine learning techniques to train the BBF-E2E system, where the mini-batch SGD algorithm is employed to search for the optimal solution to (P1).

To better illustrate the learning-based optimization process, Fig.~\ref{fig:ASMFramework} depicts the physical EM wave propagation in BBF-E2E together with its equivalent neural network expression. In the neural network expression, TX-D\textsuperscript{2}NN and RX-D\textsuperscript{2}NN serve as the encoder and decoder of the AE, respectively. Each metasurface layer is abstracted as a diffraction layer containing $N$ trainable phase parameters. The CE loss is minimized through the gradient-based training of these parameters. However, the high-dimensional matrix multiplications involved in \eqref{eq:TXNNSignalProcessing} and \eqref{eq:RXNNSignalProcessing} constitute the major computational bottleneck during training. To reduce this overhead, we introduce ASM to facilitate efficient training of large-scale diffractive neural networks. In the following subsections, we present the ASM and gradient-based training algorithms in detail.

\begin{figure*}[t]
	\centering
	\includegraphics[width=0.85\textwidth]{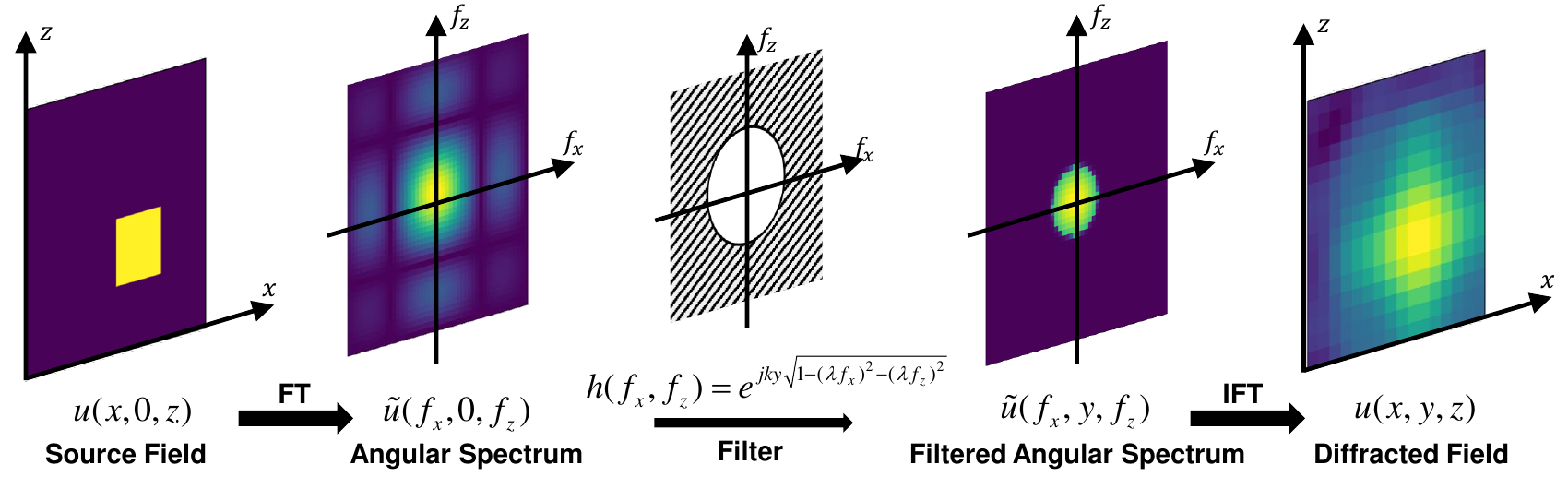}
	\vspace{-0.1cm}
	\captionsetup{justification=raggedright, singlelinecheck=false}
	\caption{The process of diffracted field calculation with ASM, which exploits the equivalence between spatial diffraction and angular spectrum filtering.}
	\label{fig:ASMMethod}
	\vspace{-0.4cm}
\end{figure*}

\subsection{Forward Propagation: The Angular Spectrum Method}
\label{s4a}
In this work, ASM is adopted instead of RSF to calculate the diffracted field. ASM exploits the equivalence between spatial diffraction and angular spectrum filtering, transforming the spatial signal into its angular spectrum for diffraction computation~\cite{goodmanIntroductionFourierOptics2005}. Based on FFT, ASM  significantly reduces the computational complexity of diffraction computation.

We begin with the fundamental equation of wave parapagation, i.e., the Helmholtz equation. It is directly derived from Maxwell’s equations in a source-free medium and governs the propagation of different kinds of waves in space. With scalar diffraction theory, we represent the EM field using a scalar function $\mathcal{U}(x, y, z, t)$, where $x$, $y$, and $z$ represent the coordinates in three-dimensional space, and $t$ is time. The Helmholtz equation is then written as:
\begin{equation}
	\nabla^2 \mathcal{U} - \frac{1}{c^2} \frac{\partial^2 \mathcal{U}}{\partial t^2} = 0,
\end{equation}
where $\nabla^2$ denotes the Laplacian, i.e., $\nabla^2 \mathcal{U} = \frac{\partial^2 \mathcal{U}}{\partial x^2} + \frac{\partial^2 \mathcal{U}}{\partial y^2} + \frac{\partial^2 \mathcal{U}}{\partial z^2}$. $c$ denote the speed of light in the transmission medium.

Assuming separability of variables, let $\mathcal{U}(x, y, z, t) = u(x, y, z) g(t)$. Then, by the separation variable method, we obtain the following equations:
\begin{equation}
	\label{eq:seperatedEquations}
	\left\{
	\begin{aligned}
		& \nabla^2 u + k^2 u = 0, \\
		& \frac{d^2 g}{dt^2} + c^2 k^2 g = 0,
	\end{aligned}
	\right.
\end{equation}
where $k$ is a separation constant. Without boundary conditions, the general plane wave solution can be obtained via further separation of variables, which is given by $u(\mathbf{r}) = c_1 e^{\pm j \mathbf{k}^\mathrm{T} \mathbf{r}}$ and $g(t) = c_2 e^{\pm j \omega t}$, where $c_1$ and $c_2$ are constants, $\mathbf{r} = [x, y, z]^\mathrm{T}$, the wave vector $\mathbf{k} = [k_x, k_y, k_z]^\mathrm{T}$ satisfies $\| \mathbf{k} \|^2 = k^2$, and $\omega = kc$. Thus we have $k = \frac{2\pi}{\lambda}$ representing the wave number with $\lambda$ denoting the wavelength.

In contrast, diffraction problems require solving the Helmholtz equation subject to boundary conditions. To address such problems, Green’s function methods can be employed, from which the widely used RSF can be derived~\cite{goodmanIntroductionFourierOptics2005}:
\begin{equation}
	u(x, y, z) = \iint_{\Sigma} u(x', 0, z') \cos \theta \left( \frac{1}{2\pi r} + \frac{1}{j\lambda} \right) \frac{e^{jkr}}{r} \, dx' \, dz',
\end{equation}
where $\Sigma$ denotes the source surface, and $\theta$ is the angle between the normal to the aperture and the line connecting the source $(x', 0, z')$ and the observation point $(x, y, z)$. Assuming a uniform field distribution over each metasurface element, this formula can be further transformed into the forms like \eqref{eq:TXNNSignalProcessing} and \eqref{eq:RXNNSignalProcessing}, which serve as a practical model for field propagation in D\textsuperscript{2}NN~\cite{anStackedIntelligentMetasurfaces2023a}. It establishes the relationship between the diffracted field at an arbitrary point and the boundary field at the source surface.


Alternatively, one can derive the ASM for diffraction calculation with lower computational complexity. Concretely, based on the coordinate system shown in Fig.~\ref{fig:systemModel}, we take the 2D Fourier transform along the $x$- and $z$-axes to the first equation in~\eqref{eq:seperatedEquations}. Define the frequency-domain function as $\tilde{u}(f_x, y, f_z) = \mathcal{F}_{x,z} \left\{ u(x, y, z) \right\}$, where $\mathcal{F} \{ \cdot \}$ denotes the Fourier transform operator. Then $\tilde{u}$ satisfies
\begin{equation}
	\frac{\partial^2 \tilde{u}}{\partial y^2} + \left[ k^2 - 4\pi^2(f_x^2 + f_z^2) \right] \tilde{u} = 0.
\end{equation}
The solution to this differential equation is
\begin{equation}
	\tilde{u}(f_x, y, f_z) = \tilde{u}(f_x, 0, f_z) \cdot e^{j k \sqrt{1 - (\lambda f_x)^2 - (\lambda f_z)^2} \, y},
\end{equation}
which describes the spectral response of the spatial frequency components $(f_x, f_z)$ after propagation over a distance $y$. The transfer function can therefore be written as
\begin{align}
	h(f_x, f_z) = e^{j k \sqrt{1 - (\lambda f_x)^2 - (\lambda f_z)^2} \, y}.
\end{align}
Taking the inverse Fourier transform gives the diffracted field at $(x, y, z)$:
\begin{equation}
	\label{eq:ASM}
	\begin{aligned}
		u(x, y, z) =& \mathcal{F}^{-1}_{f_x, f_z} \left\{ \tilde{u}(f_x, y, f_z) \right\} \\
		=& \mathcal{F}^{-1}_{f_x, f_z} \left\{ \mathcal{F}_{x,z} \left\{ u(x, 0, z) \right\} h(f_x, f_z) \right\} \\
		=& \iint_{\mathbb{R}^2} \left( \iint_{\mathbb{R}^2} u(x, 0, z) e^{-j 2\pi(f_x x + f_z z)} \, dx \, dz \right) \\ 
		&\times h(f_x, f_z) e^{j 2\pi(f_x x + f_z z)} \, df_x \, df_z.
	\end{aligned}
\end{equation}
This gives the basic concept of using ASM to calculate the diffracted field. Similar to RSF, it establishes a relationship between the boundary field and the diffracted field at any point in space. Since both RSF and ASM are derived from the Helmholtz equation under the same physical assumptions, they are theoretically equivalent~\cite{goodmanIntroductionFourierOptics2005}.

Let $k_x = 2\pi f_x$, $k_z = 2\pi f_z$, and $k_y = \sqrt{k^2 - k_x^2 - k_z^2} = k \sqrt{1 - (\lambda f_x)^2 - (\lambda f_z)^2}$, such that $k_x^2 + k_y^2 + k_z^2 = k^2$. Then, \eqref{eq:ASM} can be rewritten as
\begin{equation}
	\label{eq:ASMPlainWave}
	u(\mathbf{r}) = \iint_{\mathbb{R}^2} \tilde{u}(f_x, 0, f_z) \, e^{j \mathbf{k}^\mathrm{T} \mathbf{r}} \, df_x df_z.
\end{equation}
As discussed previously, $u(\mathbf{r}) = e^{j \mathbf{k}^\mathrm{T} \mathbf{r}}$ represents the field of a plane wave, where the direction of the wave vector $\mathbf{k}$ corresponds to the direction of propagation. Therefore, \eqref{eq:ASMPlainWave} implies that the diffracted field at any spatial location can be interpreted as a superposition of infinitely plane waves. Concretely, as illustrated in Fig.~\ref{fig:ASMMethod}, by performing a 2D Fourier transform of the boundary field $u(x, 0, z)$, we decompose $u(x, 0, z)$ into a summation of plane waves, each characterized by a tuple of spatial frequency $(f_x, f_z)$ and weighted by $\tilde{u}(f_x, 0, f_z)$. After multiplied by the transfer function $h(f_x, f_z)$ to describe propagation, the inverse Fourier transform (IFT) is used to reconstruct the diffracted field $u(x,y,z)$.

When applying ASM to calculate the wave diffraction in D\textsuperscript{2}NN, we need to discretize this method based on DFT due to the discrete nature of metasurfaces. For example, considering the wave propagation from $l$-th layer to $(l+1)$-th layer in TX-D\textsuperscript{2}NN, let $\mathbf{U}_l[n_x, n_z]$ and $\mathbf{U}_l'[n_x', n_z']$ denote the complex amplitudes of the source field at the $(n_x, n_z)$-th element of the $l$-th and the diffracted field at the $(n_x', n_z')$-th element of $(l+1)$-th layer, respectively. Their DFTs are denoted as $\bar{\mathbf{U}}_l[\bar{n}_x, \bar{n}_z]$ and $\bar{\mathbf{U}}_l'[\bar{n}_x', \bar{n}_z']$. By transforming~\eqref{eq:ASM} into its discrete form, we formulate the wave propagation between adjacent layers of TX-D\textsuperscript{2}NN as \eqref{eq:discreteASM} (at the top of the this page), where $\mathcal{DF}\{\cdot\}$ denotes the DFT operation, and $\bar{h}(\bar{n}_x, \bar{n}_z)$ is the sampled version of $h(f_x, f_z)$. 
\begin{figure*}[!t]
	\normalsize
	
	\begin{equation}
		\label{eq:discreteASM}
		\begin{aligned}
			\mathbf{U}_l'[n_x', n_z'] &= \mathcal{DF}^{-1} \left\{ \mathcal{DF} \left\{ \mathbf{U}_{l-1}[n_x, n_z] \right\} \cdot \bar{h}(\bar{n}_x, \bar{n}_z) \right\} \\
			&= \frac{1}{N_x N_z} \sum_{\bar{n}_x=1}^{N_x} \sum_{\bar{n}_z=1}^{N_z}
			\sum_{n_x=1}^{N_x} \sum_{n_z=1}^{N_z}
			\mathbf{U}_{l-1}[n_x, n_z] \, e^{j2\pi \left( \frac{(n_x' - n_x)\bar{n}_x}{N_x} + \frac{(n_z' - n_z)\bar{n}_z}{N_z} \right)} e^{j k d_\text{L} \sqrt{1 - \left( \frac{\lambda \bar{n}_x}{N_x d_x} \right)^2 - \left( \frac{\lambda \bar{n}_z}{N_z d_z} \right)^2 }}.
		\end{aligned}
	\end{equation}
	
	\hrulefill
	\vspace*{4pt}
\end{figure*}

Let $\mathbf{F}_N \in \mathbb{C}^{N \times N}$ denote the DFT matrix, where $\mathbf{F}_N[n, \bar{n}] = e^{-j 2\pi n \bar{n} / N}$. Then, the matrix form of~\eqref{eq:discreteASM} can be rewritten as
\begin{equation}
	\label{eq:ASMMatrix}
	\mathbf{U}_l' = \mathbf{F}_{N_x}^\mathrm{H} \left[ \left( \mathbf{F}_{N_x} \mathbf{U}_{l-1} \mathbf{F}_{N_z} \right) \odot \bar{\mathbf{H}} \right] \mathbf{F}_{N_z}^\mathrm{H},
\end{equation}
By Defining $\mathbf{F}_{N_x, N_z} = \mathbf{F}_{N_x} \otimes \mathbf{F}_{N_z}$ and applying the properties $\text{vec}(\mathbf{A} \mathbf{B} \mathbf{C}) = (\mathbf{C}^\mathrm{T} \otimes \mathbf{A}) \text{vec}(\mathbf{B})$ and $\text{vec}(\mathbf{A} \odot \mathbf{B})=\text{diag}(\text{vec}(\mathbf{A}))\text{vec}(\mathbf{B})$, we can further transform~\eqref{eq:ASMMatrix} into
\begin{equation}
	\mathbf{u}_l' = \mathbf{F}_{N_x, N_z}^\mathrm{H} \mathbf{\bar{H}}_\text{D} \mathbf{F}_{N_x, N_z} \mathbf{u}_{l-1},
\end{equation}
where $\mathbf{u}_l' = \text{vec}(\mathbf{U}_l')$, $\mathbf{u}_{l-1} = \text{vec}(\mathbf{U}_{l-1})$, and $\mathbf{\bar{H}}_\text{D} = \text{diag}( \text{vec}(\bar{\mathbf{H}}))$ is the diagonal form of the transfer matrix $\bar{\mathbf{H}}$. Finally, define the equivalent propagation matrix as $\mathbf{W}_\text{equ} = \mathbf{F}_{N_x, N_z}^\mathrm{H} \mathbf{\bar{H}}_\text{D} \mathbf{F}_{N_x, N_z}$, and the layer-wise diffraction formula of TX-D\textsuperscript{2}NN becomes
\begin{equation}
	\mathbf{u}_l = \mathbf{\Phi}^l \mathbf{W}_\text{equ} \mathbf{u}_{l-1},
\end{equation}
which resembles RSF but is more computationally efficient due to its FFT-compatible structure as depicted in~\eqref{eq:discreteASM}.

For RX-D\textsuperscript{2}NN, similar formulations can be derived through the same procedure,  which replace RSF to constitute the complete forward propagation algorithm of BBF-E2E. The equivalent propagation matrix is defined similarly as $\mathbf{Z}_\text{equ} = \mathbf{F}_{N_x, N_z}^\mathrm{H} \mathbf{\bar{H}}_\text{D} \mathbf{F}_{N_x, N_z}$.

\begin{remark}
	It is worth noting that the integration in \eqref{eq:ASM} or \eqref{eq:ASMPlainWave} spans all spatial frequencies $f_x$ and $f_z$, which may lead to cases where $k^2 - k_x^2 - k_z^2 \leq 0$, rendering $k_y$ purely imaginary. In this case, let $k_y = j\gamma$ with 
	\begin{align}
		\label{eq:attenuation}
		\gamma = \sqrt{-(k^2 - k_x^2 - k_z^2)} > 0,
	\end{align}
	then the transfer function becomes $h(f_x, f_z) = e^{-\gamma y}$. This indicates that the corresponding wave component decays exponentially with propagation distance, and is therefore referred to as an evanescent wave. In most practical diffraction scenarios, evanescent waves are negligible due to their rapid attenuation over distances greater than a few wavelengths. However, for D\textsuperscript{2}NN where the inter-layer spacing is sub-wavelength, evanescent waves must be accounted for.
\end{remark}

\begin{remark}
	Compared with RSF, ASM significantly reduces the computational complexity. Specifically, an implementation of the RSF via direct matrix multiplication incurs a complexity of $\mathcal{O}(N^2)$. In contrast, the angular spectrum method implemented with FFT reduces the complexity to $\mathcal{O}\big(N_x N_z \log(N_x N_z) + N_x N_z + N_x N_z \log(N_x N_z)\big) = \mathcal{O}(N \log N + N)$, while zero-padding does not affect the asymptotic order of the computational complexity. This motivates the adoption of ASM in BBF-E2E for efficient forward diffraction computation.
\end{remark}

\begin{remark}
    Due to the periodic nature of the discrete Fourier transform (DFT), directly applying discrete ASM leads to circular convolution in the spatial domain. This causes the input field to be virtually replicated across the metasurface plane, introducing non-physical interference from adjacent copies. To resolve this, zero-padding could be applied before the DFT to suppress boundary artifacts and restore equivalence to linear convolution in the central region, which ensures that diffraction propagation is accurately computed within the physical aperture of the metasurface.
\end{remark}

\subsection{Backward Propagation and Network Training}
A key distinction between BBF-E2E and conventional electronic neural networks lies in the fact that the trainable parameters of each diffraction layer in BBF-E2E are complex-valued. Conventional electronic neural networks are real-valued, where complex-valued inputs are typically split into real and imaginary parts (or magnitude and phase) to fit into real-valued computation graphs. Such a workaround is not feasible in BBF-E2E, where all forward and backward computations must remain in the complex domain to match the underlying EM wave propagation and regulation mechanism of metasurfaces. Therefore, we employ Wirtinger calculus to compute gradients in the CVNN formed by BBF-E2E~\cite{leeComplexValuedNeuralNetworks2022}.

Suppose in a specific realization, the transmitted symbol is $s_m$, so the loss function can be simplified as $\mathcal{L} = -\log p'_m$. For all $l \in \mathcal{L}_\text{RX}$ and $n \in \mathcal{N}$, the gradient of the loss with respect to the RX-D\textsuperscript{2}NN phase shift $\psi_{l,n}$ is derived as
\begin{align}
	\label{eq:lossDerivativeRX}
	\frac{ \partial \mathcal{L} }{ \partial \psi_{l,n} } =
	\left( p'_m - 1 \right) \frac{ \partial p_m }{ \partial \psi_{l,n} }
	- \sum_{\substack{\tilde{m}=1 \\ \tilde{m} \neq m}}^M p'_{\tilde{m}} \frac{ \partial p_{\tilde{m}} }{ \partial \psi_{l,n} }.
\end{align}
Define $\mathcal{N}_x(m_x) = \left\{ (m_x - 1)N_{\text{sub},x} + 1, \dotsc, m_x N_{\text{sub},x} \right\}$, $\mathcal{N}_z(m_z) = \left\{ (m_z - 1)N_{\text{sub},z} + 1, \dotsc, m_z N_{\text{sub},z} \right\}$ and $\tilde{\mathcal{N}}(m) = \left\{ ({n}_x, {n}_z) \,\middle|\, {n}_x \in \mathcal{N}_x(m_x),\, {n}_z \in \mathcal{N}_z(m_z) \right\}$ to index the region on the output plain corresponding to the $m$-th symbol, where $m = m_x M_x + m_z$. For all $m \in \mathcal{M}$, the partial derivative term in \eqref{eq:lossDerivativeRX} can be derived as
\begin{align}
	\frac{ \partial p_m }{ \partial \psi_{l,n} } =
	\sum_{(\tilde{n}_x, \tilde{n}_z) \in \tilde{\mathcal{N}}(m)}
	2 \, \mathfrak{R} \left[
	j z^{1:l}_{\tilde{n},n} x^{l+1}_{n} e^{j\psi_{l,n}} \mathbf{V}_0^*[\tilde{n}_x, \tilde{n}_z]
	\right],
\end{align}
where $\tilde{n} = \tilde{n}_x N_x + \tilde{n}_z$, $z^{1:l}_{\tilde{n},n}$ denoting the $[\tilde{n},n]$-th entry of matrix $\mathbf{Z}^{1:l} = \mathbf{Z}^1 \mathbf{\Psi}^1 \mathbf{Z}^2 \mathbf{\Psi}^2 \dots \mathbf{Z}^l$, and $x^{l+1}_n$ denoting the $n$-th entry of vector $\mathbf{x}^{l+1} = \mathbf{Z}^{l+1} \mathbf{\Psi}^{l+1} \dots \mathbf{Z}^{L_\text{RX}} \mathbf{\Psi}^{L_\text{RX}} \mathbf{v}_{L_\text{RX}}$. When ASM is used, here we have $\mathbf{Z}^{l}=\mathbf{Z}_\text{equ}$, $\forall l \in \mathcal{L}_\text{RX}$.

Similarly, the gradient with respect to the TX-D\textsuperscript{2}NN phase shift $\phi_{l,n}$ is given by
\begin{align}
	\label{eq:lossDerivativeTX}
	\frac{ \partial \mathcal{L} }{ \partial \phi_{l,n} } =
	\left( p'_m - 1 \right) \frac{ \partial p_m }{ \partial \phi_{l,n} }
	- \sum_{\substack{\tilde{m}=1 \\ \tilde{m} \neq m}}^M p'_{\tilde{m}} \frac{ \partial p_{\tilde{m}} }{ \partial \phi_{l,n} },
\end{align}
for all $l \in \mathcal{L}_\text{TX}$ and $n \in \mathcal{N}$, and the partial derivative term in \eqref{eq:lossDerivativeTX} can be further derived as
\begin{align}
	\label{eq:lastEquationofGradients}
	\frac{ \partial p_m }{ \partial \phi_{l,n} } =
	\sum_{(\tilde{n}_x, \tilde{n}_z) \in \tilde{\mathcal{N}}(m)}
	2 \, \mathfrak{R} \left[
	j y^{l+1}_{\tilde{n},n} w^{1:l}_n e^{j\phi_{l,n}} \mathbf{V}_0^*[\tilde{n}_x, \tilde{n}_z]
	\right],
\end{align}
for all $m \in \mathcal{M}$, where $y^{l+1}_{\tilde{n},n}$ is the $[\tilde{n},n]$-th entry of $\mathbf{Y}^{l+1} = \mathbf{B}_\text{RX} \mathbf{H} \mathbf{\Phi}^{L_\text{TX}} \mathbf{W}^{L_\text{TX}} \mathbf{\Phi}^{L_\text{TX}-1} \mathbf{W}^{L_\text{TX}-1} \dots \mathbf{\Phi}^{l+1} \mathbf{W}^{l+1}$ and $w^{1:l}_n$ is the $n$-th entry of $\mathbf{w}^{1:l} = \mathbf{W}^l \mathbf{\Phi}^{l-1} \mathbf{W}^{l-1} \dots \mathbf{\Phi}^1 \mathbf{W}^1 \mathbf{u}_0$. When ASM is used, here we have $\mathbf{W}^{l}=\mathbf{W}_\text{equ}$, $\forall l \in \mathcal{L}_\text{TX}$.

Based on~\eqref{eq:lossDerivativeRX}–\eqref{eq:lastEquationofGradients}, a mini-batch SGD algorithm is applied to train the BBF-E2E network. Specifically, in each batch, $N_\text{bat}$ training samples are randomly drawn from the dataset. These samples undergo forward and backward propagation and the gradients are computed through automatic differentiation, which are then used to update the phase parameters of both TX-D\textsuperscript{2}NN and RX-D\textsuperscript{2}NN according to the learning rate $\eta$:
\begin{equation}
	\begin{aligned}
		\psi_{l,n} \leftarrow \psi_{l,n} - \eta \frac{ \partial \mathcal{L} }{ \partial \psi_{l,n} }, \forall l \in \mathcal{L}_\text{RX}, \forall n \in \mathcal{N}\\
		\phi_{l,n} \leftarrow \phi_{l,n}- \eta \frac{ \partial \mathcal{L} }{ \partial \phi_{l,n} }, \forall l \in \mathcal{L}_\text{TX}, \forall n \in \mathcal{N}.
	\end{aligned}
\end{equation}
Furthermore, to enhance training stability, BN layers are incorporated into the training process, though they are not implemented in validation and test process to match the hardware constraint. The BN layer normalizes the outputs of each diffraction layer over the entire mini-batch, thereby ensuring the numerical stability during training and preventing gradient vanishing or explosion issues.

\begin{figure*}[htbp]
	\centering
	\includegraphics[width=0.95\textwidth]{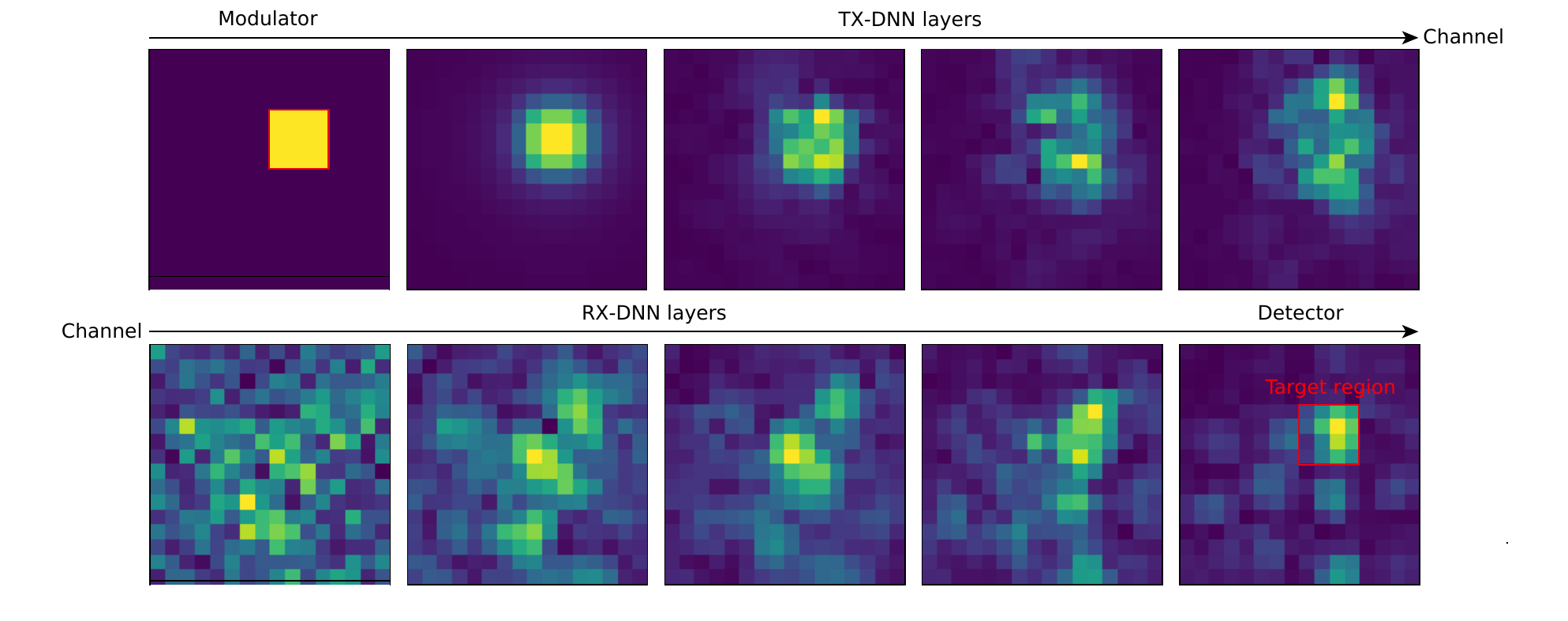}
	\vspace{-0.1cm}
	\caption{Forward propagation process of the BBF-E2E system. The heatmaps illustrate the amplitude distributions of EM waves on different layers of BBF-E2E. Initially, the modulator selects and activates a subset of antenna elements. The EM waves then propagate through the TX-D\textsuperscript{2}NN and RX-D\textsuperscript{2}NN, which are jointly optimized to mitigate the effects of channel fading and noise. Finally, the signal energy is concentrated in the target region for the detector to perform correct symbol decision.}
	\label{fig:visualization}
	\vspace{-0.4cm}
\end{figure*}

\section{Numerical Results}

\subsection{Simulation Setup}

The simulation scenario follows the setup illustrated in Fig.~\ref{fig:transceiverArchitecture} and Fig.~\ref{fig:systemModel}, where a BBF-E2E system is deployed at the transceiver to replace conventional baseband and RF modules, enabling symbol-level transmission. The operating frequency is set to 28~GHz, corresponding to a wavelength of $\lambda = 10.7$~mm. Each transmitted symbol carries 4 bits of information, i.e., $p = 4$ and $M = 16$, with $M_x=M_z=4$.

As depicted in \eqref{eq:channelModel}, a correlated Rician channel model is considered to emulate complex propagation conditions. The power normalization condition $E\{\Vert \mathbf{H} \Vert_\text{F}^2\} = N^2$ is satisfied with $E\{\Vert \mathbf{H}_\text{LoS} \Vert_\text{F}^2\} = E\{\Vert \mathbf{H}_\text{NLoS} \Vert_\text{F}^2\} = N^2$. The parameters of LoS component are set as $\theta^{\text{ele}}_\text{TX} = \theta^{\text{azi}}_\text{TX} = \theta^{\text{ele}}_\text{RX} = \theta^{\text{azi}}_\text{RX} = \frac{\pi}{4}$ without loss of generality.

The network is trained on a personal computer equipped with an Intel Core i7-12700 CPU. The network is trained by the mini-batch SGD strategy with 3200 randomly generated data samples and a batch size of $N_\text{bat} = 32$, and the Adam optimizer with a preset learning rate of 0.03 is employed to accelerate convergence.

Two primary performance metrics are used to evaluate the proposed BBF-E2E system: training loss and SER. The expression for the training loss has been given in \eqref{eq:lossFunction}, while SER is calculated by dividing the number of correctly detected symbols by the total number of transmitted symbols.

\begin{figure*}[ht!]
	\centering
	\begin{subfigure}{0.48\textwidth}
		\includegraphics[width=\linewidth]{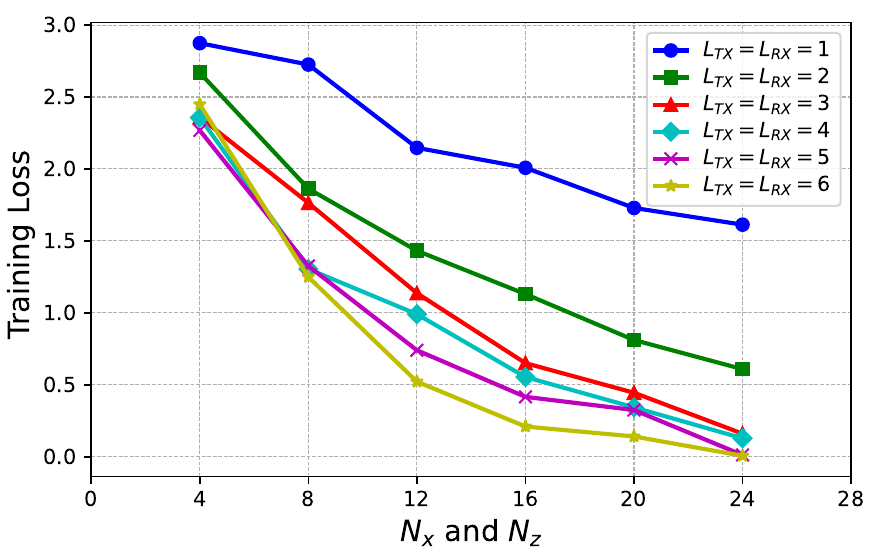}
		\caption{}
		\label{fig:sub1}
	\end{subfigure}\hfill
	\begin{subfigure}{0.48\textwidth}
		\includegraphics[width=\linewidth]{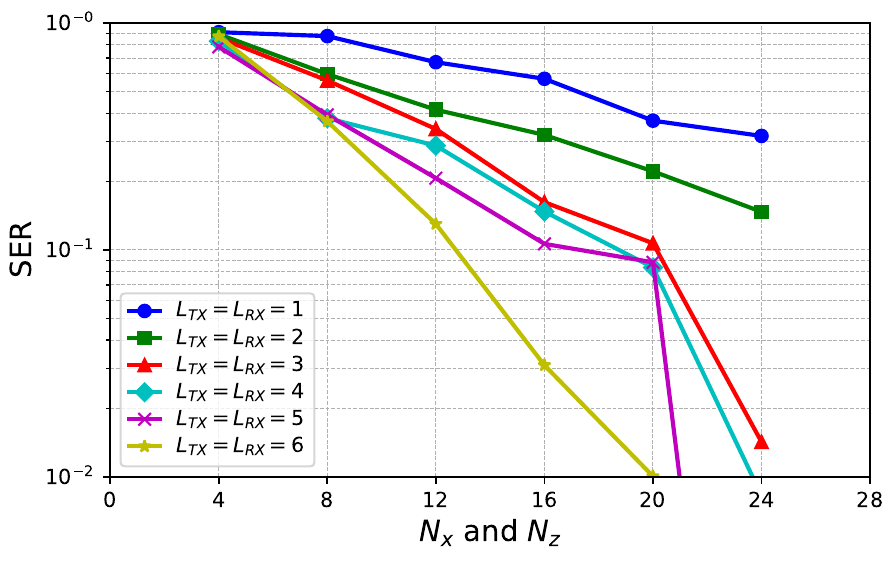}
		\caption{}
		\label{fig:sub2}
	\end{subfigure}
	\vspace{-0.1cm}
	\caption{Impact of the number of layers and elements on the performance of BBF-E2E. (a) Training loss. (b) Testing SER.}
	\label{fig:TrainingLossAndSERversusLayersandCellNum}
	\vspace{-0.4cm}
\end{figure*}

\subsection{Visualization of End-to-End Forward Propagation}
To visualize the signal propagation in the BBF-E2E system, Fig.~\ref{fig:visualization} presents an example of forward propagation process. The D\textsuperscript{2}NNs are configured with $L_\text{TX} = L_\text{RX} = 4$ layers, and each layer consists of $N = N_x \times N_z = 16 \times 16$ transmissive elements. The inter-layer spacing is set to $d_\text{L} = 1$~mm, and the element spacing is $d_x = d_z = 0.125\lambda$. The network is trained under a Rician fading channel with $K_\text{R} = 0$~dB and $\text{SNR} = -10$~dB. An input symbol is modulated by activating a selected subarray on the input plane, where only the selected antennas emit identical signals. The resulting EM wave propagates through the TX-D\textsuperscript{2}NN, traverses the noisy and faded wireless channel, and then passes through the RX-D\textsuperscript{2}NN, where the detector identifies the received pattern and recovers the transmitted symbol. Under the Rician channel condition, the EM wave transmitted by TX-D\textsuperscript{2}NN is distorted by both fading and noise, and the resulting EM wave received by RX-D\textsuperscript{2}NN becomes unrecognizable. However, after the signal processing of TX-D\textsuperscript{2}NN and RX-D\textsuperscript{2}NN, the field intensity is clearly concentrated in the target region corresponding to the originally activated area, indicating that the symbol has been reliably detected. Throughout this diffraction-based signal processing pipeline, the learned modulation and equalization scheme demonstrates strong robustness to both noise and fading. Since the EM waves in TX-D\textsuperscript{2}NN and RX-D\textsuperscript{2}NN propagate at the speed of light, the resulting system latency is extremely low, embodying the principle of ``computating-by-propagation''. By offloading computational tasks from the digital baseband to D\textsuperscript{2}NN, BBF-E2E system has the potential to drastically reduce processing latency and power consumption in future physical-layer designs.

\subsection{Training Performance with Different Model Capacities}
The number of layers and elements per layer are two critical design parameters of the BBF-E2E system, as they jointly determine the network's model capacity, which in turn impacts the system performance. Fig.~\ref{fig:TrainingLossAndSERversusLayersandCellNum} presents the trends of system performance as functions of both layer count and element count per layer. In this simulation, both training and testing SNRs are fixed at $-20$~dB. The inter-layer spacing is set to $d_\text{L} = 1$~mm, the element spacing to $d_x = d_z = 0.125\lambda$, and the Rician factor is fixed at $K_\text{R} = 0$~dB. Since the modulation order has been set to $M = 16$, $N_x$ and $N_z$ must be a multiple of $\sqrt{M} = 4$, thus we set $N_x = N_z \in \{4, 8, 12, 16, 20, 24\}$. As illustrated in Fig.~\ref{fig:TrainingLossAndSERversusLayersandCellNum}, both the training loss and testing SER decrease significantly with increasing network depth and per-layer array size. Near-zero training loss can be achieved when both network depth and per-layer array size are sufficiently large. However, when the number of elements is small (e.g., $N_x = N_z = 4$), the performance gain from adding more layers becomes marginal. The benefit of adding layers becomes more pronounced as the number of elements grows, suggesting that the trade-off between network depth and per-layer array size should be carefully balanced in system design to achieve cost-efficient hardware implementation.

\begin{figure}[t]
	\centering
	\includegraphics[width=0.48\textwidth]{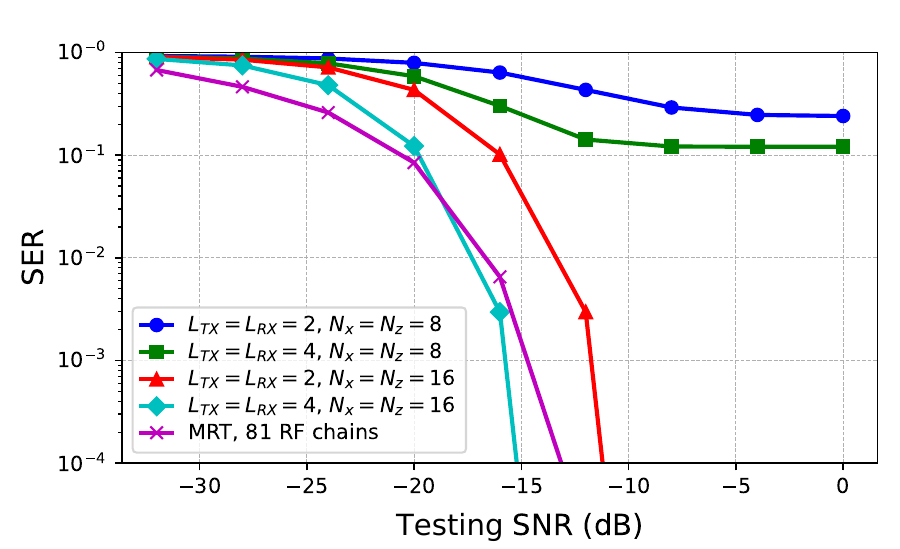}
	\vspace{-0.1cm}
	\caption{SER performance of the BBF-E2E system under different configurations, compared with a conventional transmission scheme.}
	\label{fig:SERversusSNR_DiffrentLandN}
	\vspace{-0.4cm}
\end{figure}

\begin{figure}[t]
	\centering
	\includegraphics[width=0.48\textwidth]{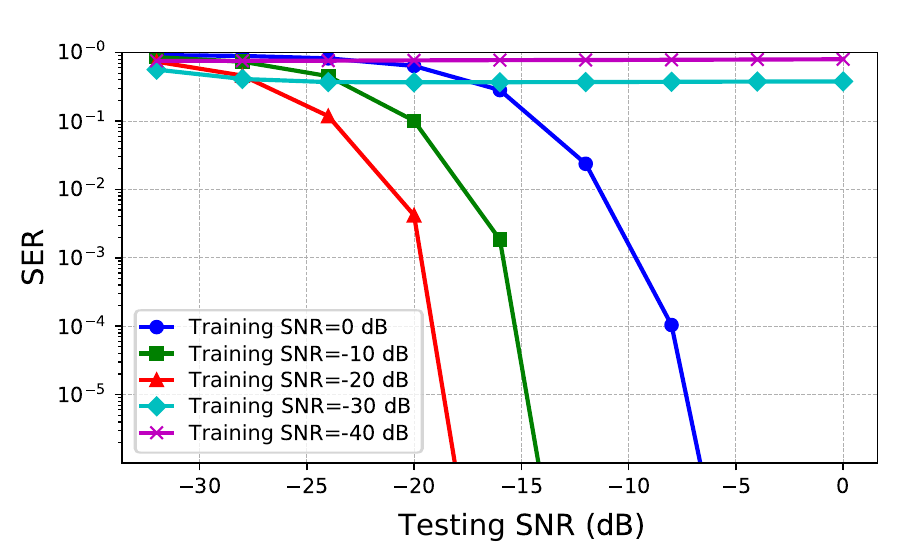}
	\vspace{-0.1cm}
	\caption{SER performance of the BBF-E2E system versus testing SNRs under different training SNRs.}
	\label{fig:SERversusSNR_DiffrentTrainingSNR}
	\vspace{-0.4cm}
\end{figure}

\begin{figure}[t]
	\centering
	\includegraphics[width=0.48\textwidth]{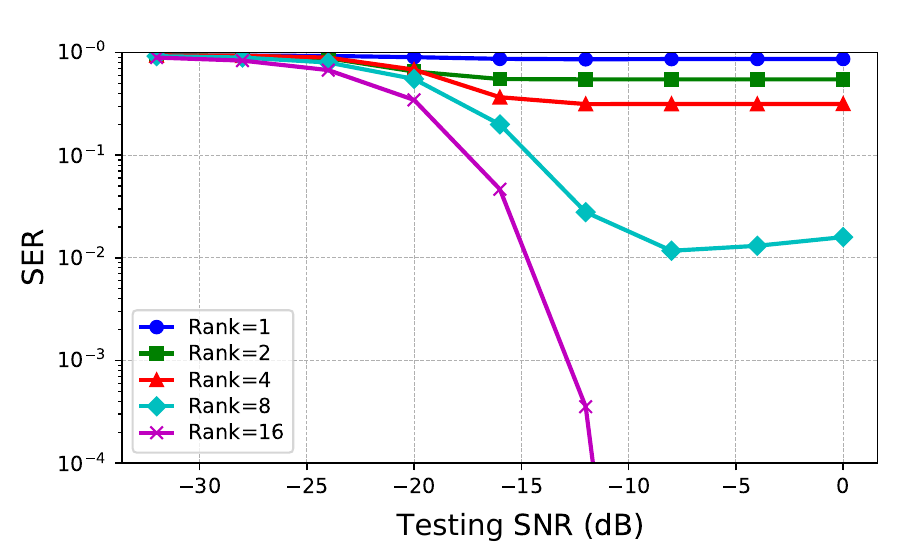}
	\vspace{-0.1cm}
	\caption{SER performance of the BBF-E2E system versus testing SNRs under different channel ranks.}
	\label{fig:Rank}
	\vspace{-0.4cm}
\end{figure}

\subsection{Testing SER Performance}

Fig.~\ref{fig:SERversusSNR_DiffrentLandN} evaluates the testing SER performance of BBF-E2E under different configurations and SNR conditions, compared with traditional modulation and beamforming techniques. Specifically, quadrature amplitude modulation (QAM) with the order of 16 and maximum ratio transmission (MRT)~\cite{loMaximumRatioTransmission1999} are selected as representatives of traditional modulation and beamforming methods. The inter-layer spacing is set to $d_\text{L} = 1$ mm, and the element spacing is $d_x = d_z = 0.125\lambda$. The BBF-E2E networks are trained under channel conditions with a Rician factor $K_\text{R} = 0$ dB and an SNR of $-10$ dB, and tested under SNR conditions ranging from $-32$ dB to $0$ dB with 4 dB intervals. About the simulation results, when $N_x = N_z = 16$, the BBF-E2E system leverages the beamforming gain provided by a large number of passive elements, achieving extremely low SER even at SNRs below $-10$ dB. However, when $N_x = N_z = 8$, the network underperforms at $\text{SNR} = -10$ dB during training, resulting in limited performance improvement despite further increases in SNR during testing. These results highlight the importance of increasing the BBF-E2E parameters for better system performance. Furthermore, when compared with the MRT transmission scheme, the BBF-E2E system with $N_x = N_z = 16$ and $L_\text{TX} = L_\text{RX} = 4$ achieves comparable performance to the MRT scheme that employs 81 RF chains. Notably, the BBF-E2E system outperforms MRT in higher SNR conditions. Regarding hardware cost, the BBF-E2E system requires only a single RF chain and 1024 passive transmissive elements, which is more cost-effective than the MRT system with 81 RF chains. This result demonstrates that BBF-E2E is a competitive low-cost solution for transmission under low SNR conditions.

Furthermore, we investigate the impact of training SNR, as shown in Fig.~\ref{fig:SERversusSNR_DiffrentTrainingSNR}. We consider a standard configuration where both the TX-D\textsuperscript{2}NN and RX-D\textsuperscript{2}NN comprise 4 layers, each consisting of 256 passive transmission elements. The inter-layer spacing is set to \( d_\text{L} = 1 \) mm, and the element spacing is \( d_x = d_z = 0.125\lambda \). The Rician factor is fixed at \( K_\text{R} = 0 \) dB. The networks are trained under various SNR conditions. Simulation results indicate that when the training SNR is extremely low (e.g., -30 dB), the network fails to learn meaningful features, resulting in poor SER performance across all testing SNRs. In contrast, training at a relatively high SNR (e.g., 0 dB) leads to better performance in high-SNR scenarios but poor performance in low-SNR scenarios. The optimal training SNR under this configuration is observed to be around -20 dB, where the network effectively learns both denoising and modulation abilities. Therefore, selecting an appropriate training SNR is crucial to achieving optimal system performance.

\subsection{Channel Impact on System Performance}
An important aspect to investigate is the impact of channel characteristics on the performance of the proposed BBF-E2E system. The following simulation reveals that the performance of BBF-E2E system exhibits a strong dependence on the channel rank.

Fig.~\ref{fig:Rank} presents the SER performance of the BBF-E2E system under different channel rank conditions. The network is trained at an SNR of $-10$ dB with configuration $L_\text{TX} = L_\text{RX} = 4$ and $N_x = N_z = 16$. The inter-layer spacing and element spacing are set to $d_\text{L} = 1$ mm and $d_x = 0.125\lambda$, respectively. The channel matrix is constructed via a rank-constrained factorization method. Specifically, two random Gaussian matrices $\mathbf{U} \in \mathbb{C}^{N \times R}$ and $\mathbf{V} \in \mathbb{C}^{R \times N}$ with rank $R$ are generated ($R\leq N$), and the channel matrix is calculated as
\begin{align}
	\mathbf{H} = \frac{N}{\Vert \mathbf{UV} \Vert _\text{F}} \mathbf{U}\mathbf{V},
\end{align}
ensuring that $\text{rank}(\mathbf{H}) \leq R$ and $\Vert \mathbf{H} \Vert_\text{F}^2 = N^2$~\cite{heathjrFoundationsMIMOCommunication2018}. The results show a clear correlation between channel rank and BBF-E2E performance. Higher-rank channels lead to better training outcomes, and the generalization ability towards higher SNR improves significantly when the channel rank achieves 16. This indicates that the BBF-E2E system naturally exploits the spatial diversity inherent in high-rank channels to enhance transmission robustness.

\begin{figure}[t]
	\centering
	\includegraphics[width=0.48\textwidth]{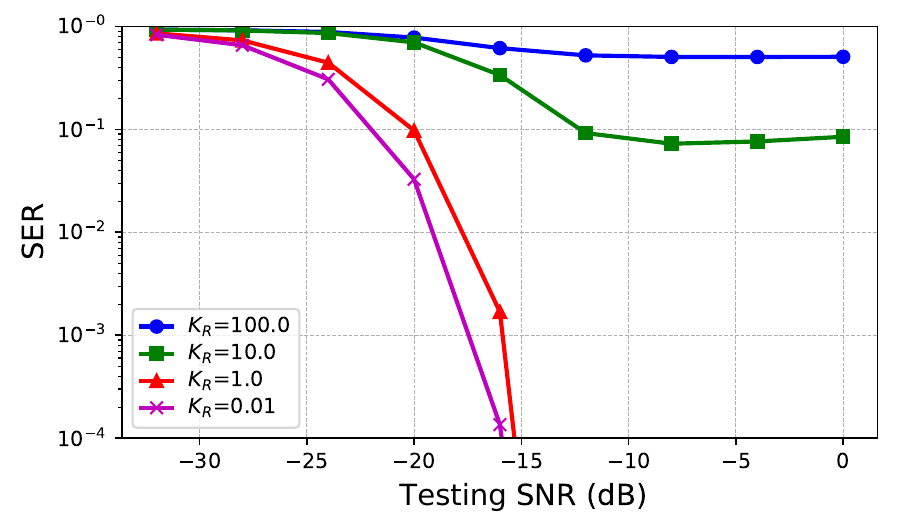}
	\vspace{-0.1cm}
	\caption{SER performance of the BBF-E2E system versus testing SNRs under different Rician factors.}
	\label{fig:RicianFactor}
	\vspace{-0.4cm}
\end{figure}

To further examine the above conclusion with physically meaningful channels, Fig.~\ref{fig:RicianFactor} investigates the performance impact of varying Rician factors under the correlated Rician fading channel, using the same system setup as in Fig.~\ref{fig:Rank}. Interestingly, the observed trends are consistent with the previous conclusion: Lower Rician factors lead to better training performance, and the system generalizes well when the factor is below or equal to 1. In contrast, larger Rician factors typically indicate less multipath component, leading to worse SER performance. These results suggest that BBF-E2E system performance is highly sensitive to the ratio of multipath component of the wireless channel.

\begin{figure}[t]
	\centering
	\includegraphics[width=0.45\textwidth]{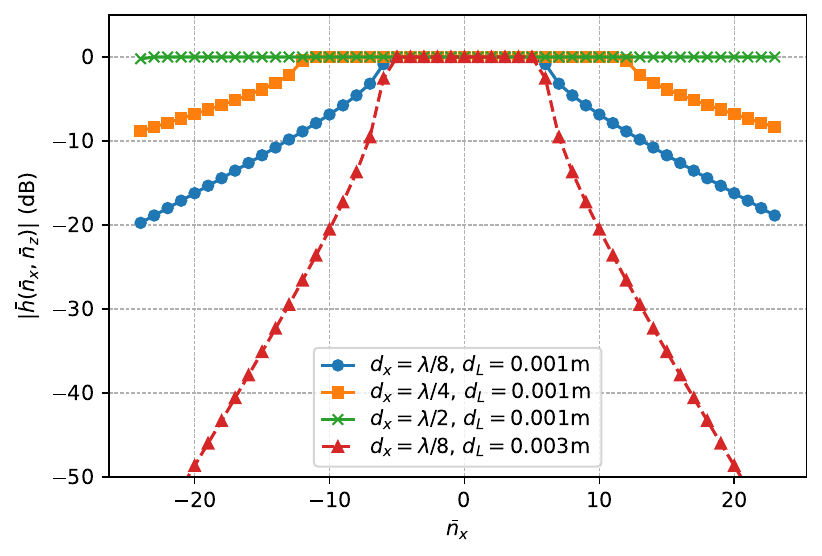}
	\vspace{-0.1cm}
	\caption{Spatial filtering effect with ASM. The amplitude of the transfer function $\bar{h}(\bar{n}_x, \bar{n}_z)$ when $\bar{n}_z=0$ is presented.}
	\label{fig:kyPlot_full}
	\vspace{-0.4cm}
\end{figure}

\subsection{Spatial filtering effect with ASM}

As illustrated in Fig.~\ref{fig:kyPlot_full}, the diffraction calculation through ASM can be understood from the perspective of ``spatial filtering'', where the discrete spatial frequency $(\bar{n}_x,\bar{n}_z)$ is divided into a passband and a stopband based on whether the amplitude of $\bar{h}(\bar{n}_x,\bar{n}_z)$ is 1 or less. The proportion of the passband and stopband is primarily influenced by the element spacing. As the element spacing increases, the transmission region expands. When $d_{x} = \frac{\lambda}{2}$, all spatial frequencies represented by $\bar{n}_x$ under the condition of $\bar{n}_z = 0$ can be transmitted without loss. Moreover, the attenuation rate within the stopband is influenced by the inter-layer spacing. This can be easily observed from \eqref{eq:attenuation}, where the larger inter-layer spacing leads to faster attenuation. In general, the ``spatial filtering'' effect of diffraction propagation offers an intuitive way to understand the ability of BBF-E2E to resist noise.

\subsection{Discussion}
The simulation results provide several key insights into the design and performance of the proposed BBF-E2E system. First, increasing the number of diffractive layers and metasurface elements consistently improves performance, confirming the value of deeper and denser wave-domain computational models. Specifically, the BBF-E2E architecture achieved performance comparable to a conventional system employing 81 RF chains, while requiring only a single RF chain and 1024 passive metasurface elements, highlighting the superior hardware efficiency of the proposed system. Even under severely degraded SNR conditions, the system reliably recovers transmitted symbols with low SER, validating its robustness. These findings suggest that the BBF-E2E architecture inherently leverages rich multipath propagation and spatial diversity, achieving strong beamforming and modulation capabilities through learned wave-domain propagation.

Looking forward, several directions remain open for future exploration. A key challenge lies in the practical hardware implementation of diffractive neural architectures, particularly in accurately fabricating and calibrating multi-layer transmissive metasurfaces. Additionally, further performance improvements can be made by exploring hybrid physical–digital architectures or hardware-in-the-loop training. Extending the BBF-E2E framework to support multi-user scenarios, semantic communications, or integrated sensing and communication (ISAC) also holds great potential for future research.

\section{Conclusion}
This work presented a novel BBF-E2E communication system that fundamentally departs from traditional baseband-dominated signal processing architectures by embedding signal modulation and beamforming into the physical propagation of EM waves. Enabled by D\textsuperscript{2}NNs, the proposed BBF-E2E system could jointly process transmitting and receiving signals via cascaded metasurface layers in an autoencoder-like manner. We established an end-to-end training framework that integrates ASM-based diffraction calculation and gradient-based network training techniques. Replacing RSF with ASF reduced the computational complexity of diffraction computing between adjacent metasurface layers from $O(N^2)$ to $O(N\log N)$. Simulation results confirmed the superiority of BBF-E2E in symbol detection accuracy and hardware efficiency, achieving comparable performance to a conventional multi-antenna system with 81 RF chains while requiring only a single RF chain and 1024 passive elements of metasurfaces. Future work may explore hardware implementation challenges, extensions to multi-user scenarios, and integration with physical–digital architectures to further exploit the potential of wave-domain neural computing in wireless systems.

\bibliographystyle{IEEEtran}
\bibliography{IEEEabrv,reference.bib}

\end{document}